\documentclass[journal]{IEEEtran}
\usepackage{amsmath}
\usepackage{amssymb}
\usepackage{amsthm}
\usepackage{cite}
\usepackage{color}
\usepackage{xcolor}
\usepackage{caption}
\usepackage{epsfig,latexsym}
\usepackage{float}
\usepackage{fancyhdr}
\usepackage{graphicx}
\usepackage{indentfirst}
\usepackage{mdwmath}
\usepackage{mdwtab}
\usepackage{subfigure}
\usepackage{setspace}
\usepackage{times}
\usepackage{url}
\usepackage{multirow}
\usepackage{algorithm}
\usepackage{algorithmic}
\usepackage{multicol}
\usepackage{subcaption}
\usepackage{amsthm,amsmath,amssymb,lipsum}
\usepackage{mathrsfs}
\usepackage{booktabs}  
\usepackage{array}     
\usepackage{booktabs}
\usepackage{tabularx} 
\newtheorem{remark}{Remark}

\makeatletter

\begin{document}
\title{\textrm{ASTARS empowered Satellite Positioning Approach for Urban Canyons and Indoor Environments}}

\author{Yu Zhang, Xin Sun, Tianwei Hou,~\IEEEmembership{Member,~IEEE,} Anna Li,~\IEEEmembership{Member,~IEEE,}  \\Sofie Pollin,~\IEEEmembership{Senior Member,~IEEE,} Yuanwei Liu,~\IEEEmembership{Fellow,~IEEE,} \\and Arumugam Nallanathan,~\IEEEmembership{Fellow,~IEEE}
\thanks{This work was supported in part by the Fundamental Research Funds for the Central Universities under Grant 2023JBZY012 and 2024JBMC014, in part by the National Natural Science Foundation for Young Scientists of China under Grant 62201028, in part by Young Elite Scientists Sponsorship Program by CAST under Grant 2022QNRC001, in part by the Beijing Natural Science Foundation L232041, and in part by the Marie Skłodowska-Curie Fellowship under Grant 101154499, in part by EPSRC grant numbers to acknowledge are EP/W004100/1, EP/W034786/1 and EP/Y037243/1.}
\thanks{Yu Zhang and Xin Sun are with the School of Electronic and Information Engineering, Beijing Jiaotong University, Beijing 100044, China (e-mail:21221191@bjtu.edu.cn; xsun@bjtu.edu.cn).}
\thanks{T. Hou is with the School of Electronic and Information Engineering, Beijing Jiaotong University, Beijing 100044, China, and also with the School of Electronic Engineering and Computer Science, Queen Mary University of London, London E1 4NS, U.K. (email: twhou@bjtu.edu.cn).}
\thanks{Anna Li is with the School of Computing and Communications, Lancaster University, Lancaster LA1 4WA, U.K. (e-mail: a.li16@lancaster.ac.uk).}
\thanks{Sofie Pollin is with the  Department of Electrical Engineering (ESAT), KU Leuve, Belgium (e-mail: sofie.pollin@esat.kuleuven.be).}
\thanks{Yuanwei Liu is with the Department of Electrical and Electronic Engineering, The University of Hong Kong, Hong Kong (e-mail: yuanwei@hku.hk).}
\thanks{Arumugam Nallanathan is with the School of Electronic Engineering and Computer Science, Queen Mary University of London, London E1 4NS, U.K., and also with the Department of Electronic Engineering, Kyung Hee University, Yongin-si, Gyeonggi-do 17104, Korea (e-mail: a.nallanathan@qmul.ac.uk).}}

\maketitle

\begin{abstract}
To mitigate the loss of satellite navigation signals in urban canyons and indoor environments, we propose an active simultaneous transmitting and reflecting reconfigurable intelligent surface (ASTARS) empowered satellite positioning approach. Deployed on building structures, ASTARS reflects navigation signals to outdoor receivers in urban canyons and transmits signals indoors to bypass obstructions, providing high-precision positioning services to receivers in non-line-of-sight (NLoS) areas. The path between ASTARS and the receiver is defined as the extended line-of-sight (ELoS) path and an improved carrier phase observation equation is derived to accommodate that. The receiver compensates for its clock bias through network time synchronization, corrects the actual signal path distance to the satellite-to-receiver distance through a distance correction algorithm, and determines its position by using the least squares (LS) method. Mathematical modeling of the errors introduced by the proposed method is conducted, followed by simulation analysis to assess their impact.
Simulation results show that: 1) in areas where GNSS signals are blocked, with time synchronization accuracy within a 10 ns error range, the proposed method provides positioning services with errors not exceeding 4 m for both indoor and outdoor receivers, outperforming conventional NLoS methods with positioning errors of more than 7 m; 2) the additional errors introduced by the proposed method do not exceed 3 m for time synchronization errors within 10 ns, which includes the phase shift, beamwidth error, time synchronization errors, and satellite distribution errors, outperforming traditional NLoS methods, which typically produce positioning errors greater than 5 m.
\end{abstract}

\begin{IEEEkeywords}
GNSS, Urban Canyon, Indoor Positioning, ASTARS.
\end{IEEEkeywords}

\section{\textrm{INTRODUCTION}}
The global navigation satellite system (GNSS) is widely acknowledged for its provision of all-weather, high-precision positioning and timing services~\cite{GNSS1}, which are extensively applied across various global applications, particularly in critical industries such as transportation, telecommunications, and agriculture~\cite{GNSS2,9262624,10197979}.
With the rapid progression of automated systems, intelligent transportation infrastructure and the forthcoming sixth generation mobile network (6G), the significance of GNSS in enhancing the accuracy and reliability of positioning systems is increasingly evident~\cite{Saad2019AVO,6Gzhanwang,YDSJ202101017}. 

Traditional GNSS relies on the line-of-sight (LoS) signals for precise positioning~\cite{8998218}.
Under unobstructed signal conditions, the receiver can achieve positioning by using navigation signals from at least four satellites. 
However, in urban canyons and indoor environments, the LoS signals from satellites are often blocked by buildings and other obstacles, which makes it difficult for GNSS receiver to acquire sufficient satellite signals, leading to a significant decline in positioning accuracy, where the positioning accuracy of satellite single-point positioning may increase to 10 m or higher~\cite{10014853}. 
As urbanization continues, these high-accurate positioning challenges in non-LoS (NLoS) environments such as the urban canyons and indoor environments have become increasingly prominent~\cite{YUAN2020106315,10431794,9840374}. 
Addressing these complex positioning issues has therefore become a critical focus in the development of GNSS technology~\cite{indoorpositioning}.
\subsection{\textrm{Related Work}}
To overcome the above issues, numerous positioning techniques have been proposed to provide solutions for urban canyons and indoor environments.
Iwase et al. proposed an urban canyon positioning technique by utilizing NLoS signals~\cite{10.1007/s10291-012-0260-1}. 
NLoS signals can provide valuable spatial geometry information that can be leveraged to improve positioning accuracy when LoS signals are blocked while they introduce errors. To effectively utilize NLoS signals while mitigating their negative effects, consistency-checking methods are developed to detect and eliminate erroneous pseudo-range observations~\cite{Groves2013HeightAC,Blanch2015FastMF}. However, the positioning error of their method is more than 5 m and in urban canyon environments, due to the high proportion of multipath signals, the performance of multipath detection and rejection is unstable.~\cite{Hsu2017MultipleFG}.
Chen et al. introduced the modified design matrix approach based on the GNSS, specifically designed to address the issue of low positioning accuracy in urban canyon environments~\cite{gaijinjuzhen}. By taking into account the effects of building reflections and obstructions on satellite signals, the modified design matrix approach reduces errors introduced by the NLoS environments to 7.14 m. However, the effectiveness of the modified design matrix approach depends on the specific urban canyon environments, particularly the distribution and material characteristics of buildings, which prevents its widespread adoption.

Yan et al. addressed the indoor satellite navigation issue by combining pseudo-lites with a navigation signal simulator~\cite{Yan2022IntersatellitePD}, in which pseudo-lites are ground-based devices that transmit signals similar to real satellites, providing auxiliary positioning in indoor environments or areas with weak satellite signals. However, establishing a pseudo-lite system requires significant hardware investment and maintenance efforts.~\cite{Gan2018IndoorCP}. 

Furthermore, researchers have also explored positioning methods that integrate GNSS with other techniques to overcome these challenges. 
Yang et al. proposed a mobile positioning technique by using signals of opportunity (SoOP) in urban canyons, which introduced a multipath dominant signal parameter estimation method that
integrates with three-dimensional (3D) city maps to achieve precise positioning within an 1-meter error range~\cite{9109876,GNSSzengqiang2}. 
However, in most urban canyon environments, obtaining accurate 3D city maps is challenging and costly.
Joseph et al. integrated GNSS with Wi-Fi signals to enhance the positioning accuracy when the GNSS signals are weak or unavailable~\cite{GNSSzengqiang1,INS,IPS}, but the quality of Wi-Fi signals is often affected by the environmental factors like building density and device location, resulting in less stable positioning with an error range typically between 5 to 15 m, which is heavily influenced by environment, making them unable to fully replace the high precision positioning of GNSS~\cite{WIFI,LEO2}.

To address these issues, we turn our focus to the reconfigurable intelligent surface (RIS), an emerging communication technology. 
RIS is a kind of planar material that can be deployed on the exterior surfaces of buildings, and its principle is similar to mirror reflection~\cite{RIS1,RIS2}. When electromagnetic signals reach RIS, it can automatically adjust its electromagnetic parameters, transforming the transmission path into a controllable form by reflecting the signals. 
Based on these characteristics, the RIS has been widely applied in various communication scenarios~\cite{9140329,9133094}.
Hou et al. proposed a communication and navigation integration scheme by using the non-orthogonal multiple access (NOMA) technique with RIS, where the communication performance was thoroughly analyzed~\cite{NOMA}.
Li et al. applied RIS to integrated navigation and communication (INAC) networks, combining RIS with satellite navigation technology, which analyzed its signal characteristics~\cite{INAC1}. 
Guan et al. experimentally demonstrated that NOMA and RIS effectively minimize the energy consumption of active beamforming, laying the groundwork for the practical application of RIS~\cite{LEOcom}. 
Zhao et al. verified through simulation that the combination of three visible satellite signals and one reflected signal from the RIS provides a positioning solution through pseudo-range measurements when the number of visible satellites is insufficient in RIS-assisted GNSS networks~\cite{INAC}. However, the method of Zhao et al. is limited to a single RIS-reflected navigation signal and three direct navigation signals. When multiple RIS-reflected signals occur, the positioning accuracy significantly degrades. Meanwhile, the ranging error caused by RIS is not systematically discussed.
\subsection{\textrm{Motivation and Contributions}}
Building on prior research in applying RIS for communication purposes, considering the low power intensity of satellite navigation signals and the diverse requirements of indoor and outdoor receivers, traditional RIS may face challenges in providing reliable positioning support under these conditions. To address these limitations, we introduce the active simultaneous transmitting and reflecting reconfigurable intelligent surface (ASTARS) to the satellite navigation. Compared to RIS, ASTARS can achieve both transmission and reflection of navigation signals, providing a reliable solution for receivers in various environments. Due to the integration of amplifier components, ASTARS can actively amplify navigation signals at the cost of some power consumption~\cite{ARIS1,ARIS2}. Additionally, the positioning error introduced by ASTARS array from phase shifts, beamwidth, and hardware delays, all of which are predictable and manageable, making ASTARS to be a reliable solution for accurate positioning.

\begin{figure}[h!]
\centering
\includegraphics[width =3in]{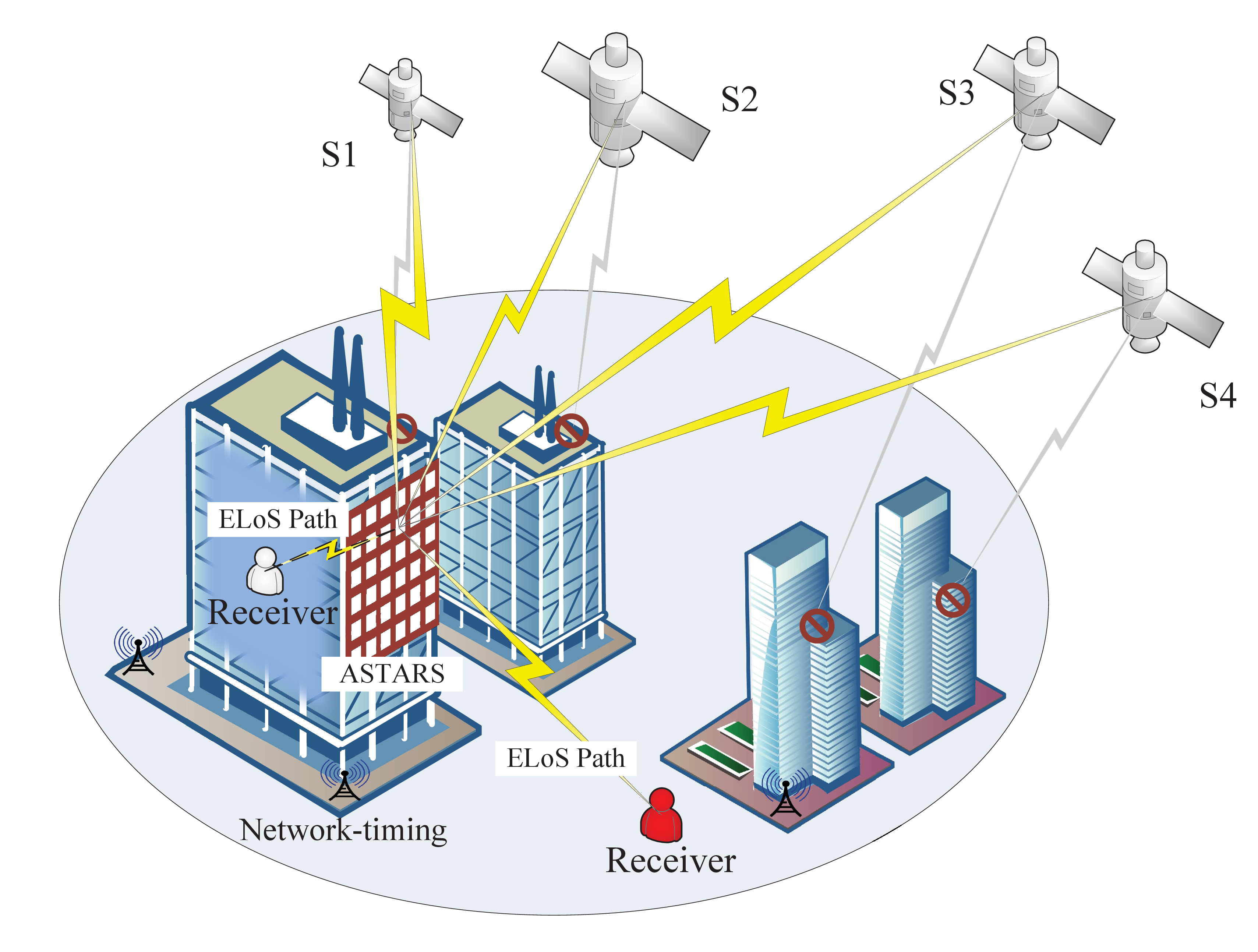}
\caption{Illustration of the proposed ASTARS empowered satellite positioning networks.}
\label{system_model}
\end{figure}

In our work, we propose the ASTARS empowered satellite positioning networks, which can provide satellite positioning services for both indoor and outdoor receivers simultaneously as shown in Fig.~\ref{system_model}. 
When the LoS signals are completely blocked, the transmission and reflection characteristics of ASTARS are utilized to create an extended LoS (ELoS) path, which refers to the path between the ASTARS and receiver.
However, the proposed ASTARS empowered satellite positioning approach poses three challenges: i) due to the existence of the ELoS path, the traditional carrier phase observation is no longer applicable to the proposed scenario; ii) the ELoS path is shared by multiple satellites, which, if not properly accounted for, can result in a shifted position estimation; iii) the receiver clock bias and the ELoS path propagation time are coupled, as both influence the signal's travel time, making it challenging to separate their effects on positioning. 

To address these challenges, we introduced a new carrier phase observation equation to accommodate the novel navigation signal transmission path. Network time synchronization technology was employed to decouple the transmission time of the ELoS path from the receiver clock bias, instead of relying on traditional GNSS algorithms to calculate the clock bias. Additionally, a geometric correction method was applied to rectify the ranging errors caused by the ELoS path. With the development of 5G, the precision requirements for time synchronization have increased. In 5G, base station (BS) requires synchronization accuracy of ±390 ns, which is tightened to ±65 ns in new radio (NR) collaborative services to ensure orthogonal frequency division multiplexing (OFDM) symbol-level alignment. In applications like indoor positioning or internet of things (IoT), the required accuracy is typically ±10 ns~\cite{timing11,timing12,timing14}. National Time Service Center, Chinese Academy of Sciences’s research has achieved ultra-high precision synchronization, reaching 5 ns, which provides the technical feasibility for the implementation of our work~\cite{CTC}.

Compared with conventional indoor positioning methods such as terrestrial pseudolites or signal repeater, the proposed ASTARS empowered satellite positioning network offers three key advantages. Firstly, it retains full compatibility with existing GNSS signals and receivers, avoiding the need for customized hardware or new signal structures. Secondly, it leverages available network infrastructure, such as 6G BS, to provide precise timing references, which are increasingly accessible in urban environments. Thirdly, ASTARS is cost-effective and energy-efficient, which has a wide coverage area, providing a highly efficient and scalable solution for enhancing positioning performance in GNSS-denied scenarios.

The main contributions of this paper are as follows:
\begin{itemize}
    \item We develop a novel ASTARS empowered satellite positioning approach, which provides high-precision positioning services to receivers in urban canyons and indoor environments simultaneously by using signals transmitted and reflected by ASTARS. 
    \item We propose an ELoS path model, and then investigate the impact of transmission and reflection channels on navigation signals via the ELoS path. The network time synchronization is integrated with the satellite positioning approach to calculate the ELoS path distance. We propose a new carrier phase observation model based on the ELoS path to accommodate the novel navigation signal transmission path, which estimates the ASTARS position. 
    \item We conduct an in-depth analysis of error sources in the proposed ASTARS empowered satellite positioning approach. Through theoretical analysis and simulation validation on phase shift error, beamwidth error of ASTARS, time synchronization error and satellite distribution, we establish an accurate and reliable satellite positioning framework, which provides a solid foundation for enhancing the overall performance and reliability of satellite positioning approach.
    \item We validate the proposed approach in the simulated urban canyons and indoor environments. The experimental results show that: 1) when the ASTARS can detect more than 5 satellites while the network time synchronization error is controlled within ±10 ns, the positioning error for urban canyon and indoor receivers is within 4 m, which outperforms the positioning accuracy of over 5 m achieved through the NLoS method; 2) when the network time synchronization error is controlled within ±10 ns, the newly introduced errors of the proposed approach is within 3 m. 
\end{itemize}

\subsection{\textrm{Organization and Notations}}
This paper is organized as follows: Section \uppercase\expandafter{\romannumeral2}  introduces the ASTARS channel, signal models, and the ELoS path-based positioning model. Section \uppercase\expandafter{\romannumeral3} discusses the network time synchronization and the receiver positioning algorithms. Section \uppercase\expandafter{\romannumeral4} analyzes the newly introduced errors. Section \uppercase\expandafter{\romannumeral5} provides simulation analyses of error components and positioning results. The final section summarizes the work.
\begin{table}[h!]
\centering
\caption{ Major notations summary.}
\resizebox{0.48\textwidth}{!}{%
\begin{tabular}{lc}
\toprule

		$K$ & number of ASTARS elements \\
        $E$ & transmission scenario\\
        $R$ & reflection scenario\\
        $\psi(t)$ & incident navigation signal at time $t$\\
        $\psi _k^E(t)$ & transmitted navigation signal at time $t$\\
        $\psi _k^R(t)$ & reflected navigation signal at time $t$\\
        $E_k$ & transmission coefficient\\
        $R_k$ & reflection coefficient\\
        $H$ & amplification factor\\
        $c$ & speed of light\\
        $\lambda$ & wavelength of the navigation signal\\
		$r_{i,Rs}$ & distance between the $i$-th satellite and ASTARS\\
		$r_{Ru}$      & distance between ASTARS and the receiver\\
		$r_{i,su}$    & distance between the $i$-th satellite and the receiver\\
        $\tilde r_{i,su}$    & computed distance between the $i$-th satellite and the receiver\\
        ${\tilde\varphi _i}$ &carrier phase observation\\
		${\varphi}_i$     & simplified carrier phase observation  \\
		$T_u$     & receiver clock bias  \\
		$T^i$   & clock bias of the $i$-th satellite\\
		$T_R$   & ELoS path propagation time\\
        $\hat T_R$ & estimated ELoS path propagation time\\
		$(x_R,y_R,z_R)$ & receiver position\\
		$(x_u,y_u,z_u)$   &  ASTARS position\\
		$(x^i,y^i,z^i)$ &  position of the $i$-th satellite\\
		$(x_0,y_0,z_0)$   & initial Taylor expansion point\\
		$(x_n,y_n,z_n)$   &  $n$-th least squares iterative solution\\
        $\varepsilon$ &  unmodeled errors in carrier phase observation\\
        $\omega$ & error introduced by ASTARS\\
        $\gamma$ & network time synchronization error\\
        $\rho _i$ & distance calculation error\\
\bottomrule
\end{tabular}%
}
\label{error-analysis}
\end{table}
\section{\textrm{SYSTEM MODEL}}
This section introduces the basic concept of ASTARS and the valuable parameters. It also describes the constructed transmission and reflection channel models. An improved carrier phase observation model based on the ELoS path is also proposed.
\subsection{ \textrm{Channel Model and Signal Propagation Model}}
The ASTARS is made from the metamaterials, which are two-dimensional (2D) material structures with programmable macroscopic physical properties. Its most significant feature is a reconfigurable electromagnetic wave response. In the proposed ASTARS empowered satellite positioning
network, it can control the channel between the satellite and receiver, converting and radiating waves into the desired propagating waves in free space.

Wireless signals are electromagnetic waves propagating in 3D space. For the ASTARS, the Love's field equivalence principle states that the electromagnetic field inside and outside a closely packed surface can be uniquely determined by the currents and magnetic fields on the surface~\cite{Love}. 
Assuming that the ASTARS array has $K$ elements, within each element, the intensity and distribution of these equivalent currents are determined by the incident signals ${\psi (t)}$, as well as the local surface average electric impedance ${Z_k}$ and magnetic impedance ${Y_k}$. Assuming that the transmitted and reflected signals generated by the ASTARS have the same polarization, these signals at the $k$-th element can be represented as:

\begin{equation}\label{STARRIS_1}
\begin{aligned}
\psi _k^E(t) = {E_k}{\psi (t)},\\
{\text{  }}\psi _k^R(t) = {R_k}{\psi (t)},
\end{aligned}
\end{equation}
where ${E_k}$ and ${R_k}$ represent the transmission and reflection complex coefficients of the $k$-th element, respectively. 
According to the law of energy conservation~\cite{1999}, for ASTARS element, the following constraints on the local transmission and reflection coefficients must be satisfied:
\begin{equation}\label{STARRIS_2}
{\left| {{E_k}} \right|^2} + {\left| {{R_k}} \right|^2} \leqslant H,
\end{equation}
where $H$ is the amplification factor. It is worth noting that due to the presence of the amplifier, the amplification factor will be greater than 1 and will vary depending on the performance of the amplifier.

From the perspective of the ASTARS design, supporting the magnetic currents is the key to achieve independent control of both the transmitted and reflected signals. By incorporating equivalent surface electric and magnetic currents into the model, which characterizes varying surface impedances over the elements and time, the proposed hardware model can independently characterize the transmission and reflection properties of each element.

For the convenience of designing ASTARS in wireless communication systems, these narrow band frequency-flat coefficients are rewritten in terms of their amplitude and phase shift as follows:
\begin{equation}\label{STARRIS_4}
\begin{aligned}
{E_k} = \sqrt{\beta _k^E}{e^{j\theta _k^E}},\\{\text{  }}{R_k} = \sqrt{\beta _k^R}{e^{j\theta _k^R}},
\end{aligned}
\end{equation}
where $\beta _k^E$ and $\beta_k^R$ are real-valued coefficients satisfying $\beta _k^E+\beta_k^R \leqslant H$. $\theta_k^E$ and $\theta_k^R$ are the phase shifts introduced by the $k$-th element for the transmitted and reflected signals, respectively.

The channel vector $\boldsymbol{h}^{i}$ between the $i$-th satellite and ASTARS elements is expressed as:
\begin{equation}\label{STARRIS_5}
 \boldsymbol{h}^{i}= \left[ {\begin{array}{*{20}{c}}
  {{h_1^{i}}}&{{h_2^{i}}}& \cdots &{{h_k^{i}}} & \cdots &{{h_K^{i}}}
\end{array}} \right]^T,
\end{equation}
where $h_i^k$ represents the channel response between the $i$-th satellite and the $k$-th ASTARS element.
Because that the satellite-to-ASTARS link is dominated by a clear LoS component while retaining low-power diffuse reflections, the large-scale fading magnitude $|h_i^k|$ is well described by a Rician distribution \cite{9140329}.

The transmitted channel matrix $\boldsymbol{g}^{E}$ between the ASTARS and indoor receiver is:
\begin{equation}\label{STARRIS_6}
 \boldsymbol{g}^{E}= \left[ {\begin{array}{*{20}{c}}
  {g_1^{E}}&{{g_2^{E}}}& \cdots &{g_k^{E}}& \cdots &{g_K^{E}} 
\end{array}} \right]^T,
\end{equation}
and the reflected channel matrix $\boldsymbol{g}^{R}$ between the ASTARS and receiver in urban canyons is:
\begin{equation}\label{STARRIS_6}
 \boldsymbol{g}^{R}= \left[ {\begin{array}{*{20}{c}}
  {g_1^{R}}&{{g_2^{R}}}& \cdots &{g_k^{R}}& \cdots &{g_K^{R}} 
\end{array}} \right]^T,
\end{equation}
where $g_{k}^{E}$ and $g_{k}^{R}$ represent the transmitted channel response and the reflected channel response between the $k$-th ASTARS element and receiver, respectively. Since the ELoS path is close to the ground and subject to varying degrees of obstruction, the small-scale fading $|g_{k}^{E}|$ and $|g_{k}^{R}|$ follow the Rayleigh distribution \cite{9140329}.

Therefore, the signal $\zeta (t)^{i}$ from the $i$-th satellite via ASTARS is expressed as:
\begin{equation}\label{STARRIS_7}
\zeta (t)^{i} = (({\boldsymbol{g}^{A}})^{T}{\boldsymbol{R}^A}{\boldsymbol{h}^{i}} r_{i,Rs}^{ - \frac{{{\alpha _1}}}{2}}r_{A,Ru}^{ - \frac{{{\alpha _2}}}{2}})\sqrt P \psi (t) + {N_0},
\end{equation}
where ${\boldsymbol{R}^A}$ denotes the diagonal matrix with ${\boldsymbol{R}^A}=diag[\begin{array}{*{20}{c}}
  { {\beta _1^A} e^{j\theta _1^A}}&{ {\beta _2^A} e^{j\theta _2^A}}& \cdots &{ {\beta _k^A} e^{j\theta _k^A}} 
\end{array}]$, $A \in \left\{E, R\right\}$, $r_{i,Rs}$ represents the distance between the $i$-th satellite and ASTARS, $r_{A,Ru}$ represents the distance of the ELoS path. ${\alpha _1}$ and  ${\alpha _2}$ denote the pass loss exponent of the satellite-ASTARS link and the ASTARS-receiver link, $r_{i,Rs}^{ - \frac{{\alpha _1}}{2}}$ and $r_{A,Ru}^{ - \frac{{\alpha _2}}{2}}$ represent the large-scale fading components of the signal. $P$ denotes the transmit power of navigation signal. ${N_0}$ denotes the additive white Gaussian noise (AWGN).
\subsection{\textrm{Principle and Application of ASTARS System}}

The ASTARS system changes the incident GNSS wavefront by applying a programmable phase shift on its $K$ elements, with the navigation data, PRN code, and carrier remaining unchanged. 
Because that there is no base-band demodulation, duplication, or retransmission, ASTARS behaves as a wavefront sculptor rather than an active repeater, thereby inherently avoiding self-jamming or harmful interference~\cite{RIS1}.

%

When navigation signals impinge on the ASTARS array, the different path lengths from the satellite to each element create a deterministic phase gradient across the surface. 
The ASTARS controller captures a short snapshot of the complex baseband signals at the $K$ elements and performs lightweight signal discrimination by using frequency filtering and coarse code correlation to isolate navigation signal components. 
Subsequently, it constructs the spatial covariance matrix and applies the multiple-signal-classification (MUSIC) super-resolution algorithm to estimate the angle of arrival (AoA), denoted as \({\theta}_{\text{AoA}}\)~\cite{1143830}. 
Once \({\theta}_{\text{AoA}}\) is obtained, the desired angle of departure (AoD), \(\theta_{\text{AoD}}\), is derived based on the LoS geometry between the array centroid and the intended receiver.

During the initial access phase, the ASTARS controller in time-division-duplex systems acquires a coarse estimate of the receiver’s position by processing uplink pilot signals transmitted by the receiver. Leveraging channel reciprocity, the controller estimates the AoD of the signal observed in the uplink channel, which enables the controller to configure the phase weights of the ASTARS elements to form a narrow beam directed toward the estimated AoD to facilitate highly directional signal transmission~\cite{DOA}.
Furthermore, ASTARS supports real-time tracking of low-dynamic receivers by exploiting AoD sensing and beam tracking, enabling sub-nanosecond beam alignment~\cite{beamtrack}. The resulting additional ranging error is less than 1 mm, which is well below typical GNSS positioning noise. Consequently, the hardware delay introduced by ASTARS can be considered negligible in the overall error budget.

Beyond wavefront manipulation, ASTARS is designed to support diverse requirements in 6G-integrated space-air-ground networks by operating across multiple frequency bands for both communication and navigation. Since higher-frequency communication signals require denser element spacing to maintain beamforming accuracy, while lower-frequency navigation signals benefit from wider spacing to avoid mutual coupling and preserve radiation efficiency, ASTARS adapts by selectively activating element subsets to emulate the desired spacing, despite its physically fixed array structure.
In terms of deployment, ASTARS can be mounted on existing infrastructure such as communication towers, satellite antennas, or building rooftops. For navigation-centric applications, it is also well suited for integration into glass curtain walls, advertising panels, or other urban surfaces offering favorable LoS conditions and wide-area coverage. Depending on the application and coverage needs, the physical footprint of an ASTARS array typically spans several square meters but can exceed 10 square meters in special cases.

\begin{remark}
It is important to note that the AoA and AoD parameters used in our method are integral to the functioning of ASTARS, which are treated as known quantities within the system.
\end{remark}
\begin{remark}\label{position} 
A low-power ($\leq$ 10 dBm) 2.4 GHz beacon transmits the AoA/AoD data every 0.1 s. The total rate is $\leq$ 0.5 kbps and the link is completely outside the GNSS bands. The message is used only for ELoS path correction, which does not create extra pseudorange or carrier-phase observables, and the ASTARS position is still solved by the receiver from its own carrier-phase measurements.
\end{remark}

\subsection{\textrm{Positioning Model}}
 Assuming that the receiver continuously tracks and locks onto the satellite signals to obtain absolute carrier phase observations. According to~\cite{HofmannWellenhof1992GlobalPS}, the direct carrier phase observation $\tilde\varphi _i$ between the $i$-th navigation satellite and receiver at the observation epoch can be described by:
\begin{equation}\label{1}
\begin{aligned}
  {\tilde\varphi _i}\lambda  =&{r_{i,su}} - c{T_u} + cT^i - {N_i}\lambda  \hfill \\
   &- {V_{i,ion}} - {V_{i,trop}} + \varepsilon, \hfill \\ 
\end{aligned}
\end{equation}
where ${ \tilde\varphi}_i$ denotes the carrier phase observations obtained by the receiver (in radians), $\lambda$ is the wavelength of satellite signals (in meters), ${r_{i,su}}$ represents the geometric distance between the receiver and the $i$-th satellite (in meters), $c$ is the speed of light in vacuum (in meters per second), $T_u$ is the receiver clock bias (in seconds), $T^i$ is the satellite clock bias (in seconds), $N_i$ is the integer ambiguity (in cycles), ${V_{i,ion}}$ and ${V_{i,trop}}$ are the ionospheric delay and the tropospheric delay, respectively (in meters), and $\varepsilon$ represents the unmodeled errors including measurement noise and multipath effect (in meters).
Receiver measurement noise typically follows a zero-mean Gaussian distribution $\mathcal{N}(0, {\sigma_{n}}^2)$, where the variance ${\sigma_{n}}^2$ depends on the signal-to-noise ratio (SNR) and receiver characteristics~\cite{wucha3}.
Multipath errors exhibit spatial-temporal correlations and non-Gaussian characteristics in harsh environments, but can be approximated as white Gaussian noise for static receivers with proper antenna design and site selection~\cite{wucha1}.
Therefore, the $\varepsilon$ can be modeled as a Gaussian process for filter implementation.

$T^i$ can be resolved by using the accurate mathematical models and correction products from organizations like the international GNSS service (IGS). $N_i$ can be solved by using the LAMBDA algorithm~\cite{Teunissen1995}. ${V_{i,ion}}$  and ${V_{i,trop}}$ are corrected in the observations by utilizing well-known models such as the Klobuchar model for ionospheric delay and the Saastomoinen model for tropospheric delay~\cite{4104345,Saastamoinen2013AtmosphericCF}. 
By compensating the modeled errors and delays, and ignoring the unmodeled errors, the carrier phase observation equation can be simplified to:
\begin{equation}\label{model}
    {\varphi}_i \lambda  = {r_{i,su}} - c{T_u}+ \varepsilon.
\end{equation}

\begin{figure}[h!]
\centering
\includegraphics[width =2.3in]{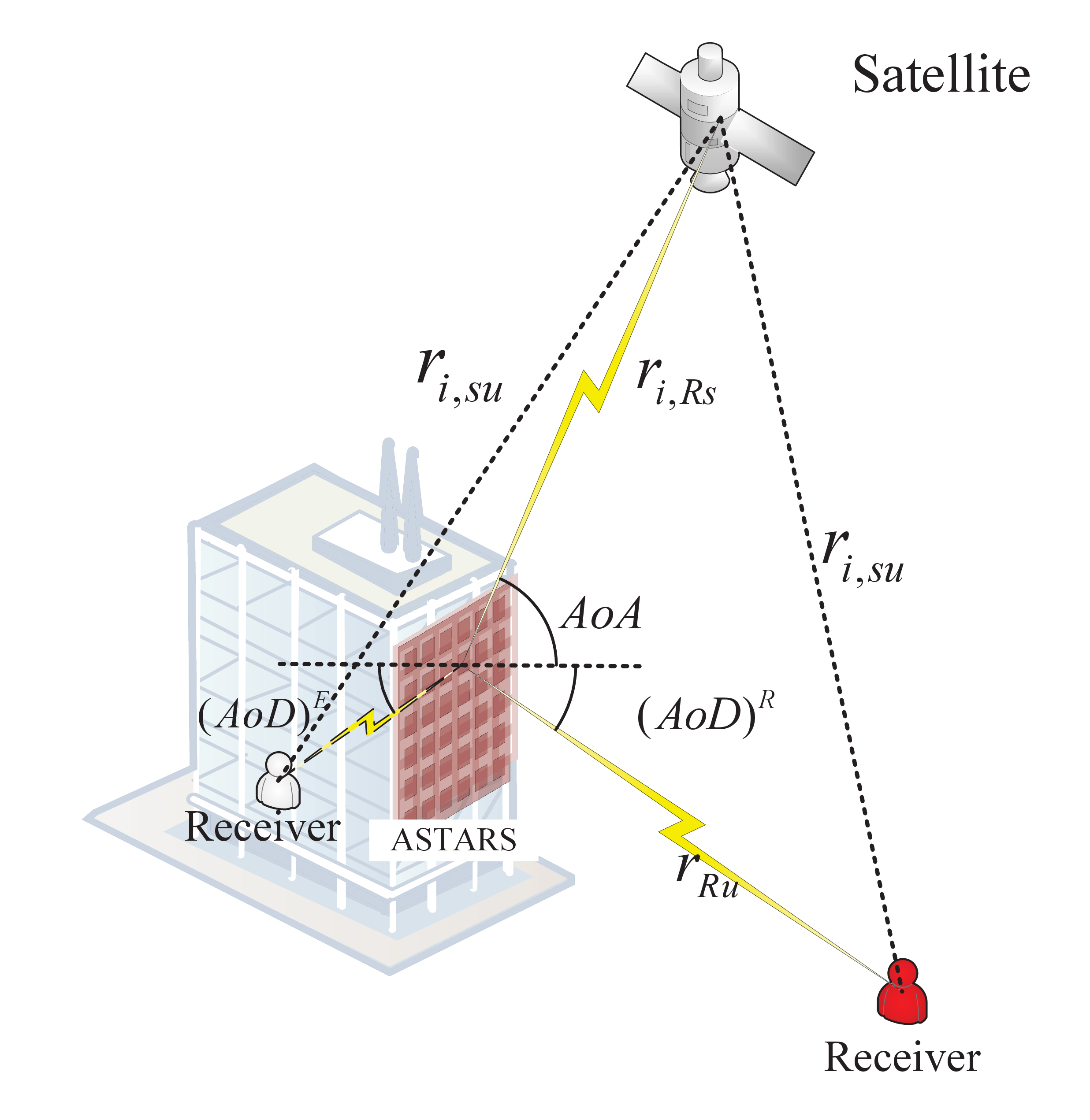}
\caption{Schematic diagram of the transmission and reflection paths and angles of ASTARS.}
\label{angel}
\end{figure}

According to Fig.~\ref{angel}, in the proposed ASTARS empowered satellite positioning networks, the propagation distance changes from ${r _{i,su}}$ to ${r _{i,Rs}} + r _{Ru}+\omega$. $\omega $ donates the distance error introduced by the ASTARS, and will be analyzed in {\bf{Section \uppercase\expandafter{\romannumeral4}}}. Based on the physical model of signal propagation aided by ASTARS, (\ref{model}) can be rewritten to
\begin{equation}\label{model_xiuzheng_1}
{\varphi}_i \lambda  = {r_{i,Rs}} + {r_{Ru}} - c{T_u}+ \varepsilon+\omega .
\end{equation}

The distance of the ELoS path can be expressed as:
\begin{equation}\label{zhuanhuan1}
r_{Ru}=c{T_R},
\end{equation}
 where $T_R$ is the time of flight from ASTARS to the receiver. 
 
Based on (\ref{zhuanhuan1}), combining the receiver clock bias and the propagation time of the signal from ASTARS to the receiver, (\ref{model_xiuzheng_1}) can be rewritten to
\begin{equation}\label{model_xiuzheng_2}
{\varphi}_i \lambda  = {r_{i,Rs}} - c({T_u}-{T_R})+ \varepsilon+\omega.
\end{equation}

\section{\textrm{ASTARS Empowered Positioning Approach}}
This section introduces the ASTARS empowered satellite positioning algorithm, including four key components: ASTARS position calculation algorithm, ELoS path distance estimation, satellite-to-receiver distance correction algorithm, and the receiver positioning algorithm.

\subsection{\textrm{ASTARS Position Calculation Algorithm}}
By observing signals from $i$ satellites, we can derive the following carrier phase observation equations as follows:
\begin{equation}\label{carrier_phase}
\left\{ {\begin{array}{*{20}{c}}
{{\varphi _1}\lambda  = {r_{1,Rs}} - c({T_u} - {T_R}) + {N_1}\lambda  + \varepsilon  + \omega, }\\
{{\varphi _2}\lambda  = {r_{2,Rs}} - c({T_u} - {T_R}) + {N_2}\lambda  + \varepsilon  + \omega, }\\
 \cdots \\
{{\varphi _i}\lambda  = {r_{i,Rs}} - c({T_u} - {T_R}) + {N_i}\lambda  + \varepsilon  + \omega. }
\end{array}} \right.
\end{equation}

Through the process as shown in \textbf{Algorithm~\ref{alg:dd_ar_en}}, the ambiguity can be calculated and inverted back to the carrier phase equation set by the RTK DD-AR approach in the following algorithm~\cite{LAMBDA}.

\begin{algorithm}[h]
\small
\caption{RTK Double-Difference Ambiguity Resolution (DD-AR)}
\label{alg:dd_ar_en}
\begin{algorithmic}[1]
  \REQUIRE Synchronous carrier-phase observations $\Phi_i^{(\cdot)}$ from rover $U$ and base $B$;
           precise/broadcast ephemerides \& clocks; wavelength $\lambda$;
           satellite unit vectors $\mathbf e_i\;(i=1,\dots,m)$
  \ENSURE  Centimetre-level baseline $\mathbf b=[\Delta X,\Delta Y,\Delta Z]^{\mathrm T}$;
           fixed DD integers $\mathbf n=[N_{21}^{UB},\dots,N_{m1}^{UB}]^{\mathrm T}$;
           undifferenced rover integers $N_i^{U}$

  \STATE \textbf{Form single differences:}
         $\Delta\Phi_i^{UB} \leftarrow \Phi_i^{U}-\Phi_i^{B}$
  \STATE \textbf{Form double differences} (satellite~1 reference):
         $\nabla\!\Delta\Phi_{i1}^{UB} \leftarrow \Delta\Phi_i^{UB}-\Delta\Phi_1^{UB}$,
         $i=2,\dots,m$
  \STATE \textbf{Linearise geometry:}
         $\nabla\!\Delta\rho_{i1}^{UB}\approx \mathbf e_{i1}^{\mathrm T}\mathbf b$
  \STATE \textbf{Least-squares estimation} $\rightarrow$
         float solutions $\hat{\mathbf b}_f,\;\hat{\mathbf n}_f$
  \STATE \textbf{LAMBDA}: decorrelate $\rightarrow$ integer search $\rightarrow$ ratio test
  \IF{ratio test accepted}
      \STATE Obtain fixed DD integers $\mathbf n^{\mathrm{fixed}}$
      \STATE \textbf{Compute fixed baseline:}
             $\mathbf b_{\mathrm{fixed}}
             \leftarrow \hat{\mathbf b}_f-\mathbf Q_{bn}\mathbf Q_{nn}^{-1}
                       (\hat{\mathbf n}_f-\mathbf n^{\mathrm{fixed}})$
  \ELSE
      \STATE Accumulate more data or switch reference satellite
      \STATE \textbf{goto} Step~3
  \ENDIF
  \medskip
  \STATE \textbf{Remarks:}
  \STATE \hspace{0.5em}DD integers $N_{i1}^{UB}\in\mathbb Z$ $(i=2,\dots,m)$
  \STATE \hspace{0.5em}SD integer $\Delta N_{1}^{UB}=N_1^{U}-N_1^{B}\in\mathbb Z$
  \STATE \hspace{0.5em}Base-station PPP-AR integers $N_i^{B}\in\mathbb Z$
  \STATE \hspace{0.5em}Undifferenced rover integers
         $N_i^{U}=N_{i1}^{UB}+\Delta N_{1}^{UB}+N_i^{B},\; i=2,\dots,m$
\end{algorithmic}
\end{algorithm}

Expanding the position $({x_0},{y_0},{z_0})$ around the $(x_R,y_R,z_R)$ by using a Taylor series, the linearized measurement equation is obtained as:
\begin{equation}\label{9}
\begin{aligned}
  \Delta {r _i}  \approx  &\frac{{{x}^{i}-{x_0} }}{{{r}^{i}}}\Delta x+\frac{{{y}^{i}-{y_0} }}{{{r}^{i}}}\Delta y \hfill \\
   &+ \frac{{{z}^{i}-{z_0} }}{{{r}^{i}}}\Delta z + c\Delta T, \hfill \\ 
\end{aligned}
\end{equation}
where $\Delta {r _i}= {{\varphi }_i}\lambda  - {r}^{i}$ represents the residual between the carrier phase observation value received by the receiver and the calculated value at the Taylor expansion point, ${r}^{i}$ is the geometric distance between the $i$-th satellite and Taylor expansion point, $\Delta x = {x_R} - {x_0}$, $\Delta y = {y_R} - {y_0}$, $\Delta z = {z_R} - {z_0}$, and $T=T_u-T_R$. In order to simplify the form of presentation and to facilitate calculations, (\ref{9}) can be rewritten into a compact form as:
\begin{equation}\label{11}
\Delta {\mathbf{r}} = {\mathbf{A}} \times {{\mathbf{B}}},
\end{equation}
where 
\begin{equation}\label{12}
\begin{aligned}
\begin{gathered}
  \Delta {\mathbf{r}} = {\left[ {\begin{array}{*{20}{c}}
  {\Delta {r_1}}&{\Delta {r_2}}& \cdots &{\Delta {r_i}} 
\end{array}} \right]^T}, \hfill \\
  {\mathbf{B}} = {\left[ {\begin{array}{*{20}{c}}
  {\Delta x}&{\Delta y}&{\Delta z}&{\Delta T} 
\end{array}} \right]^T}, \hfill \\
  {\bf{A}} = \left[ {\begin{array}{*{20}{c}}
{\frac{{{x}^{1} - {x_0}}}{{{r}^{1}}}}&{\frac{{{y}^{1} - {y_0}}}{{{r}^{1}}}}&{\frac{{{z}^{1} - {z_0}}}{{{r}^{1}}}}&c\\
{\frac{{{x}^{{2}} - {x_0}}}{{{r}^{2}}}}&{\frac{{{y}^{2} - {y_0}}}{{{r}^{2}}}}&{\frac{{{z}^{2} - {z_0}}}{{{r}^{2}}}}&c\\
 \vdots & \vdots & \vdots & \vdots \\
{\frac{{{x}^{i} - {x_0}}}{{{r}^{i}}}}&{\frac{{{y}^{i} - {y_0}}}{{{r}^{i}}}}&{\frac{{{z}^{i} - {z_0}}}{{{r}^{i}}}}&c
\end{array}} \right].
\end{gathered}
\end{aligned}
\end{equation}

Clearly, there are four unknown parameters in (\ref{12}). When the receiver can detect signals transmitted from at least four satellites, the equation is resolvable. Thus, the ordinary LS solution to (\ref{11}) is given by 
\begin{equation}\label{15}
{\bf{{\mathbf{\hat B}}  = (}}{{\bf{A}}^{\bf{T}}}{\bf{A}}{{\bf{)}}^{{\bf{ - 1}}}}{{\bf{A}}^{\bf{T}}}{\bf{\Delta\mathbf{r} }}.
\end{equation}

The $n$-th iteration solution is:
\begin{equation}\label{LSdiedai}
\left[ {\begin{array}{*{20}{c}}
  {{x_n}} \\ 
  {{y_n}} \\ 
  {{z_n}} \\ 
  {{T_n}} 
\end{array}} \right] = \left[ {\begin{array}{*{20}{c}}
  {{x_{n - 1}}} \\ 
  {{y_{n - 1}}} \\ 
  {{z_{n - 1}}} \\ 
  {{T_{n - 1}}} 
\end{array}} \right] + \left[ {\begin{array}{*{20}{c}}
  {\Delta x} \\ 
  {\Delta y} \\ 
  {\Delta z} \\ 
  {\Delta T} 
\end{array}} \right].
\end{equation}

Following (\ref{15}), the variation vector $\mathbf{\hat B}$ is determined and added to the initial estimate to obtain an updated iterative value. The updated value is then iteratively refined until the norm $\|\mathbf{\hat B}\|$ falls below a predefined accuracy threshold. 
When the convergence criterion is satisfied, the $n$-th resulting vector 
$[x_n,y_n,z_n]$
represents the 3D coordinates of the ASTARS.

\begin{remark}\label{position} 
ASTARS operates as a collective system of multiple elements, transmitting an equivalent navigation signal to the receiver. Since the receiver perceives this signal as the result of the combined action of all elements, we define the position of ASTARS as the geometric center (centroid) of the array, which represents the overall spatial effect of ASTARS, accounting for the collective influence of its elements. The AoA and AoD are measured relative to this geometric center. 
\end{remark}

\subsection{\textrm{ELoS Path Distance Estimation}}
Through the iterative convergence of (\ref{LSdiedai}), the 3D coordinates of the ASTARS and the joint estimate of $T_R$ and $T_u$ are obtained. Because that the ELoS path manifests itself at the receiver as an additional signal propagation delay, the ELoS path propagation time $T_R$ is intrinsically coupled with the receiver clock bias $T_u$, making them difficult to separate. To overcome the above limitation, this subsection introduces network time synchronization to externally constrain the receiver clock bias. By utilizing high-precision time synchronization protocols and network timing services available in infrastructure-rich 5G and 6G environments, nanosecond-level clock bias accuracy can be achieved~\cite{SATKA2023102852}. Accordingly, the joint temporal term $T_R + T_u$, which is estimated through the LS solution in (\ref{LSdiedai}), can be regarded as the ELoS propagation time contaminated by a residual error, where the influence of network synchronization error is analyzed in detail in \textbf{Section~IV}.

\begin{figure}[h!]
\centering
\includegraphics[width =2.5in]{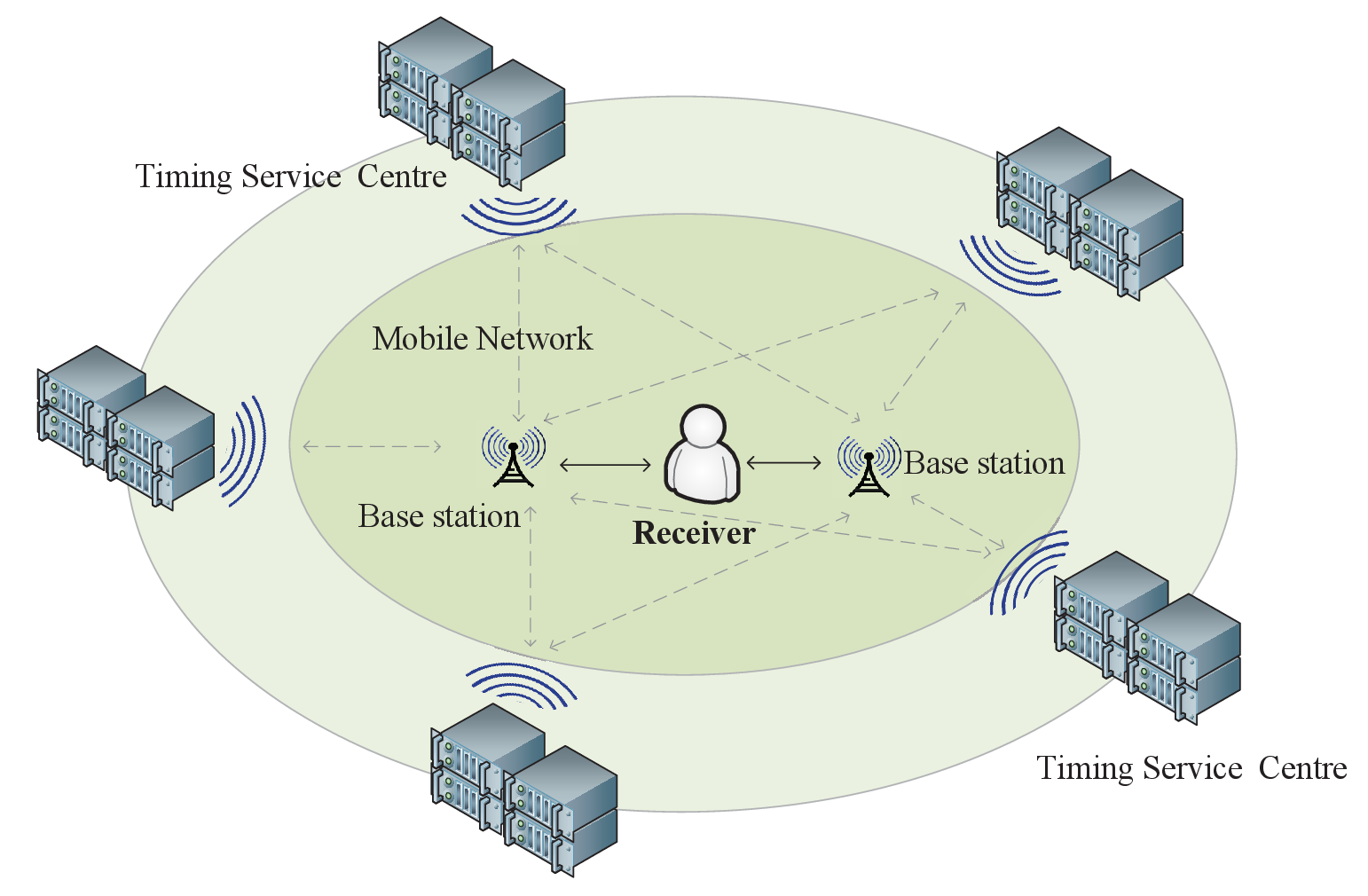}
\caption{ Illustration of the network time synchronization.}
\label{network_timing}
\end{figure}

As shown in Fig.~\ref{network_timing}, the receiver accesses the mobile networks enabling ultra-high precision synchronization with the clock from the timing service center, where the receiver clock bias is controlled within the timing accuracy range, thus the timing ofset in iteration n $T_{n}$ in (\ref{LSdiedai}) can be similar to:
\begin{equation}\label{yuedengyu} 
 T_{n}\approx -{\hat T_R}+{\gamma },
\end{equation}
where ${\hat T_R}$ denotes the computed signal transmission time of the ELoS path and ${\gamma}$ represents the network time synchronization error (in seconds). Therefore, the ELoS path distance can easily be calculated as:
\begin{equation}\label{dis_RIS_re} 
{r_{Ru}} = c{\hat T_R}.
\end{equation}

\subsection{\textrm{Distance Correction and Receiver Positioning Algorithm}}
After obtaining the position of ASTARS and the ELoS path distance, this subsection utilizes the acquired information, along with the AoA and AoD provided by ASTARS, to implement satellite-receiver path correction and achieve receiver positioning.

The ASTARS is capable of calculating the AoA $\theta_{AoA}$ and AoD $\theta_{AoD}$ as shown in Fig.~\ref{angel}. While transmitting and reflecting signals, the ASTARS broadcasts the angle information to the receiver simultaneously. Thus, the receiver can calculate the geometric angle ${\alpha _i^A}$ of the ELoS path and the corresponding cosine value ${\chi _i}$ of this angle as follows:
\begin{subequations}\label{angle}
\begin{align}
    {\alpha _i^A}  &= {\theta_{AoA}} + {(\theta_{AoD})_i^R},A=R\label{angle_1}\\
    {\alpha _i^A}  &= \pi-{\theta_{AoA}} + {(\theta_{AoD})_i^E},A=E\label{angle_1}\\
    {\chi _i} &= \cos {\alpha _i^A}.\label{angle_2}
\end{align}
\end{subequations}
As shown in Fig.~\ref{angel}, the geometric distance between the satellite and receiver can be calculated by solving the triangle formed by the satellite, ASTARS, as well as receiver. Combining (\ref{angle_2}) and using the cosine rule of the triangle, the distance between the satellite and receiver can be obtained as follows:
\begin{equation}\label{cosine rule} 
\tilde r_{i,su} = \sqrt {{{(r_{i,Rs})}^2} + {r_{Ru}}^2 - 2r_{i,Rs}{r_{Ru}}{\chi _i}}.
\end{equation}

The distance between the satellite and receiver in (\ref{cosine rule}) can be transformed into the coordinate form as follows:
\begin{equation}\label{model_diatance}
  r_{i,su} = \sqrt {{{({x^{i}} - {x_u})}^2} + {{({y^{i}} - {y_u})}^2} + {{({z^{i}} - {z_u})}^2}}. 
\end{equation}

The geometric distance equation is obtained by synthesizing the receiver continuous reception of signals from multiple satellites as follows :
\begin{equation}\label{group_diatance} 
\left\{ \begin{gathered}
  {\tilde r_{1,su} = r_{1,su}}+{\rho _1},  \\
  \begin{array}{*{20}{c}}
  {\tilde r_{2,su} = r_{2,su}}+{\rho _2}, \\ 
   \cdots  \\ 
  {\tilde r_{i,su} = r_{i,su}}+{\rho _i},
\end{array} \hfill \\ 
\end{gathered}  \right.
\end{equation}
where $\rho_i$ (in meters) represents the calculation error, primarily caused by quantization errors and AoA angle estimation errors.
The AoA estimation error $\varepsilon_{\text{AoA},i}$ arises from the MUSIC algorithm. For a single source observed by an $K$-element half-wavelength uniform linear array with $L$ snapshots and linear SNR $\rho_{s}$, which is well approximated by a zero-mean Gaussian law:
\begin{equation}
\varepsilon_{\text{AoA},i}\sim\mathcal{N}\!\!(0,\sigma_{\text{AoA}}^{2}),
\end{equation}
where
$\sigma_{\text{AoA}}^{2}\!\approx\!6\big/\!(\rho_{s}\,L\,K\,(K^{2}-1))$ (in radians$^{2}$).  
The AoD term is deterministic because the transmit beam is steered to a preset angle, whose error will be integrated into \textbf{Section IV.B} for analysis.
The quantisation error $q_i$ (in meters) results from encoding the commanded phase/angle with finite resolution~$\Delta$, which follows a uniform distribution
$q_i\sim\mathcal{U}(-\tfrac{\Delta}{2},\tfrac{\Delta}{2})$
with variance $\sigma_{q}^{2}=\Delta^{2}/12$. 
 
By using the Taylor series expansion for (\ref{model_diatance}) based on the similar steps from (\ref{9}) to (\ref{15}), the 3D coordinates of the ASTARS obtained from (\ref{LSdiedai}) are used as initial values. By employing the LS algorithm for iterative calculations, the process continues until the positioning results converge to a predefined threshold, at which point the 3D coordinates of the receiver are determined. 

The flowchart of the proposed ASTARS empowered satellite positioning approach is shown in Fig.4.
The satellite carrier signal is transmitted or reflected by ASTARS to the receiver, while ASTARS simultaneously computes and broadcasts the corresponding AoA and AoD. Leveraging network time synchronization, the receiver performs the LS-1 fit on the raw carrier-phase observations to obtain the 3D coordinates of ASTARS and the ELoS path propagation time. By using these results together with the AoA and AoD, the receiver applies a path-correction algorithm to derive the geometric satellite to receiver ranges. The LS-2 fit is then carried out with the corrected ranges, solving the final receiver position.
\begin{figure}[h!]
\centering
\includegraphics[width =3.3in]{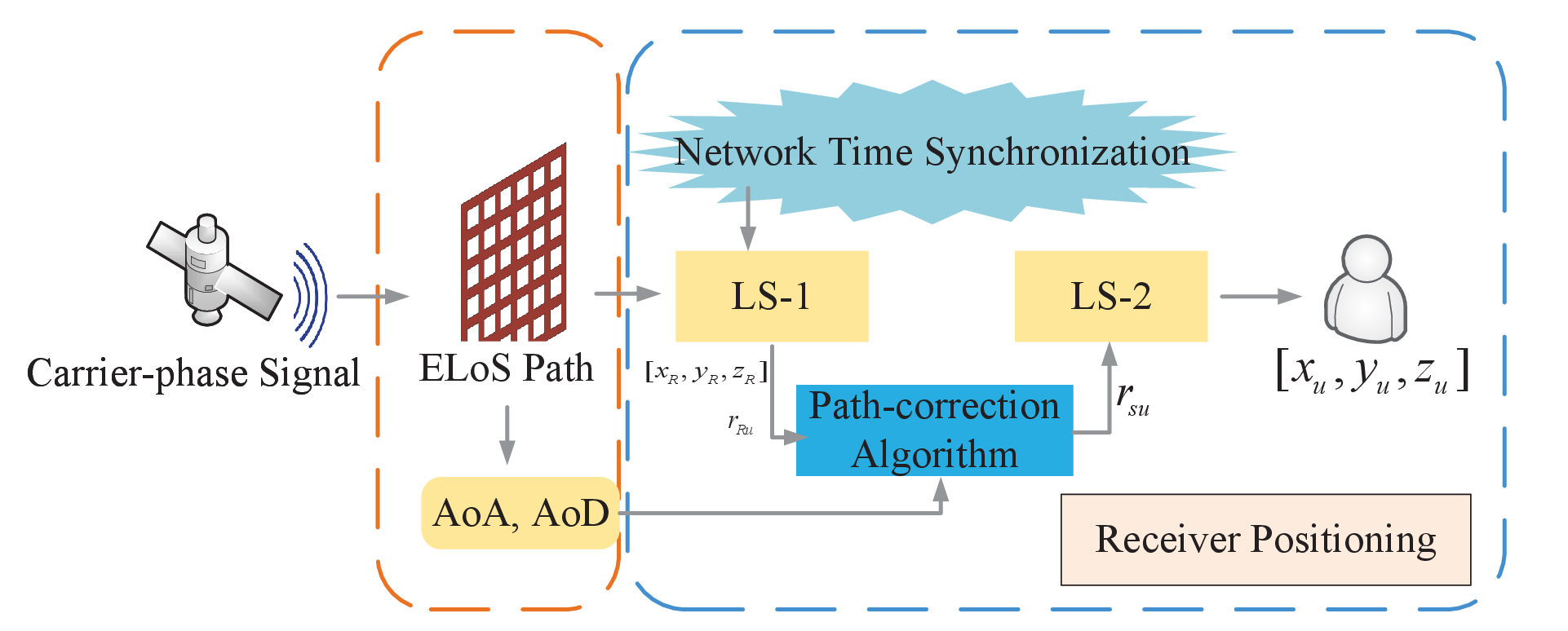}
\caption{ Flowchart of the proposed ASTARS empowered satellite positioning approach.}
\label{flowchart}
\end{figure}

\section{\textrm{ERROR ANALYSIS}}
In this section, we delve into the impact of errors introduced by ASTARS on positioning, which includes phase shift errors, beamwidth errors, time synchronization errors, and satellite distribution errors. Finally, we provide an overall expression for these errors.
\subsection{\textrm{Phase Shift Error}}
The phase shift $ \theta_k^A $ introduced by ASTARS in (\ref{STARRIS_4}) directly impacts the measurement of the carrier phase compared to an ideal specular reflector, which can be expressed as:
\begin{equation}\label{Phase Shift}
\theta_k^A = \theta_{\text{hardware}} + \theta_{\text{amp}} + \theta_{\text{noise}},
\end{equation}
where $ \theta_{\text{hardware}} $ is the hardware delay phase shift introduced as the signal propagates through the ASTARS, $ \theta_{\text{amp}} $ is the phase shift that is introduced by the amplifier during the signal amplification process, $ \theta_{\text{noise}} $ represents the small phase shift introduced by system noise, which is typically related to the SNR.
The phase shift $ \theta_k^A $ introduced by ASTARS is typically within the range of 0 to $2\pi$ and if $\theta_k^A$ exceeds $2\pi$, the integer part will be absorbed by the integer ambiguity. Since ASTARS only adjusts the propagation direction of the navigation signal through phase shift, which can be realized within a phase cycle. Therefore, the overall phase shift of the signal theoretically only produces an error of 1{-}3 wavelengths, and when ASTARS adjusts the beam according to the receiver's position, the phase drift is continuous and the rate is much lower than the bandwidth of the receiver's carrier phase-locked loop. Thus, although the phase shift may accumulate to more than $2\pi$, the change is smooth and will not trigger carrier phase cycle slips.

\subsection{\textrm{Beamwidth Error}}
Since the number of ASTARS elements is finite, resulting in a finite beamwidth, the receiver may be located anywhere within this beam. The beamwidth-induced uncertainty in the receiver position is a significant error source. 
Beamwidth is typically characterized by the half-power beamwidth (HPBW), which is defined as the angular width between the directions where the main lobe's power drops to half of its maximum value. For the ASTARS array, the beam angular width ${\theta _{beam}}$ can be approximately expressed as:
\begin{equation}\label{HPBW}
{\theta _{beam}}\approx \frac{{2\lambda }}{eL},
\end{equation}
where $L$ is the spacing among the ASTARS elements, $e$ represents the number of ASTARS elements per row.

Let us define that $\mathbf{B}_0$ is the directional vector pointing to the beam center, and $\mathbf{B}$ is the actual directional vector of the receiver. The radial distance covered by the beam $\Delta b$ (in meters) can be approximately expressed as:
\begin{equation}\label{3c}
\Delta b \approx {r_{Ru}}\tan (\frac{{{\theta _{beam}}}}{2})={r_{Ru}}\tan(\frac{\lambda}{eL}).
\end{equation}

Consequently, the beamwidth-induced uncertainty in the receiver position $\Delta p$ can be expressed as:
\begin{equation}\label{3p}
\Delta p \approx \Delta b = {r_{Ru}}\tan(\frac{\lambda}{eL}).
\end{equation}

Under the assumption of a uniform distribution, the probability density function (PDF) of the positioning error is given by:
\begin{equation}\label{3pdf}
{f_{\Delta p}}(x) = \frac{1}{{\Delta {p}}}.
\end{equation}

Based on~(\ref{3pdf}), the expectation of the positioning error $\mathbb{E}[\Delta p]$ can be expressed as:
\begin{equation}\label{3e}
\mathbb{E}\left[ {\Delta p} \right] = \int_0^{\Delta {p}} {x{f_{\Delta p}}(x)dx}=\frac{{r_{Ru}}\tan(\frac{\lambda}{eL})}{2} .
\end{equation}

\subsection{\textrm{Network Time Synchronization Error}}
Time synchronization accuracy impacts the measurement of the receiver clock bias, typically achieving nanosecond-level precision in the 6G networks. Network time synchronization error primarily arise from two sources: network delay and clock drift.

The time synchronization error $\gamma$ can be estimated by:
\begin{equation}\label{time1}
\gamma \approx \text{Max}(\delta) + \frac{\upsilon ^2}{f},
\end{equation}
where $\delta$ is the delay variation, $\upsilon $ is the standard deviation of measurements, and $f$ is the number of measurements per second.

Thus, the network time synchronization error $\Delta d$ is calculated as:
\begin{equation}\label{time3}
\Delta d \approx c \cdot \gamma.
\end{equation}

\subsection{\textrm{Satellite Distribution and Positioning Error Analysis} }
The geometric distribution of satellites directly impacts the dilution of precision (DoP), which is a critical indicator of the effect of satellite distribution on positioning accuracy. The DoP value can be calculated by using the following formula:
\begin{equation}\label{Q}
\mathbf{Q} = {({\mathbf{G}^T}\mathbf{G})^{ - 1}},
\end{equation}
where $\mathbf{G}$ is the observation matrix, defined as:
${\mathbf{G}} = \left[ {\begin{array}{*{20}{c}}
  {\frac{{x^{1} - {x_u}}}{{{r_{1,su}}}}}&{\frac{{y^{1} - {y_u}}}{{{r_{1,su}}}}}&{\frac{{z^{1} - {z_u}}}{{{r_{1,su}}}}}&1 \\ 
  {\frac{{x^{2} - {x_u}}}{{{r_{2,su}}}}}&{\frac{{y^{2} - {y_u}}}{{{r_{2,su}}}}}&{\frac{{z^{2} - {z_u}}}{{{r_{2,su}}}}}&1 \\ 
   \vdots & \vdots & \vdots & \vdots  \\ 
  {\frac{{x^{i} - {x_u}}}{{r_{i,su}}}}&{\frac{{y^{i} - {y_u}}}{{r_{i,su}}}}&{\frac{{z^{i} - {z_u}}}{{r_{i,su}}}}&1 
\end{array}} \right]$.

The specific components of the DoP value can be extracted from the diagonal elements from $\mathbf{G}$, including position DoP (PDoP), horizontal DoP (HDoP), and vertical DoP (VDoP):
\begin{equation}\label{XDOP}
\begin{gathered}
  PDoP = \sqrt {{q_{11}} + {q_{22}} + {q_{33}}},  \hfill \\
  HDoP = \sqrt {{q_{11}} + {q_{22}}},  \hfill \\
  VDoP = \sqrt {{q_{33}}},  \hfill \\ 
\end{gathered} 
\end{equation}
where $q_{ii}$ represents the  diagonal elements of G, with $i \in \left\{1,2,3 \right\}$. PDoP indicates the overall accuracy of the positioning solution, which depends on the distribution of satellite signals across both azimuth and elevation angles. HDoP measures the accuracy of positioning in the horizontal plane (X and Y axes), which depends on the distribution of satellite signals across different azimuth angles, with a uniform spread of signals from the east, west, north, and south directions being crucial for accurate horizontal positioning. VDoP assesses the accuracy in the vertical direction (Z axis), which is primarily influenced by the distribution of satellite signals in terms of elevation angle, with high-elevation satellites being crucial for improving vertical positioning accuracy.

\begin{figure}[h!]
  \centering
  \subfigure[]{{\includegraphics[width =2.8in]{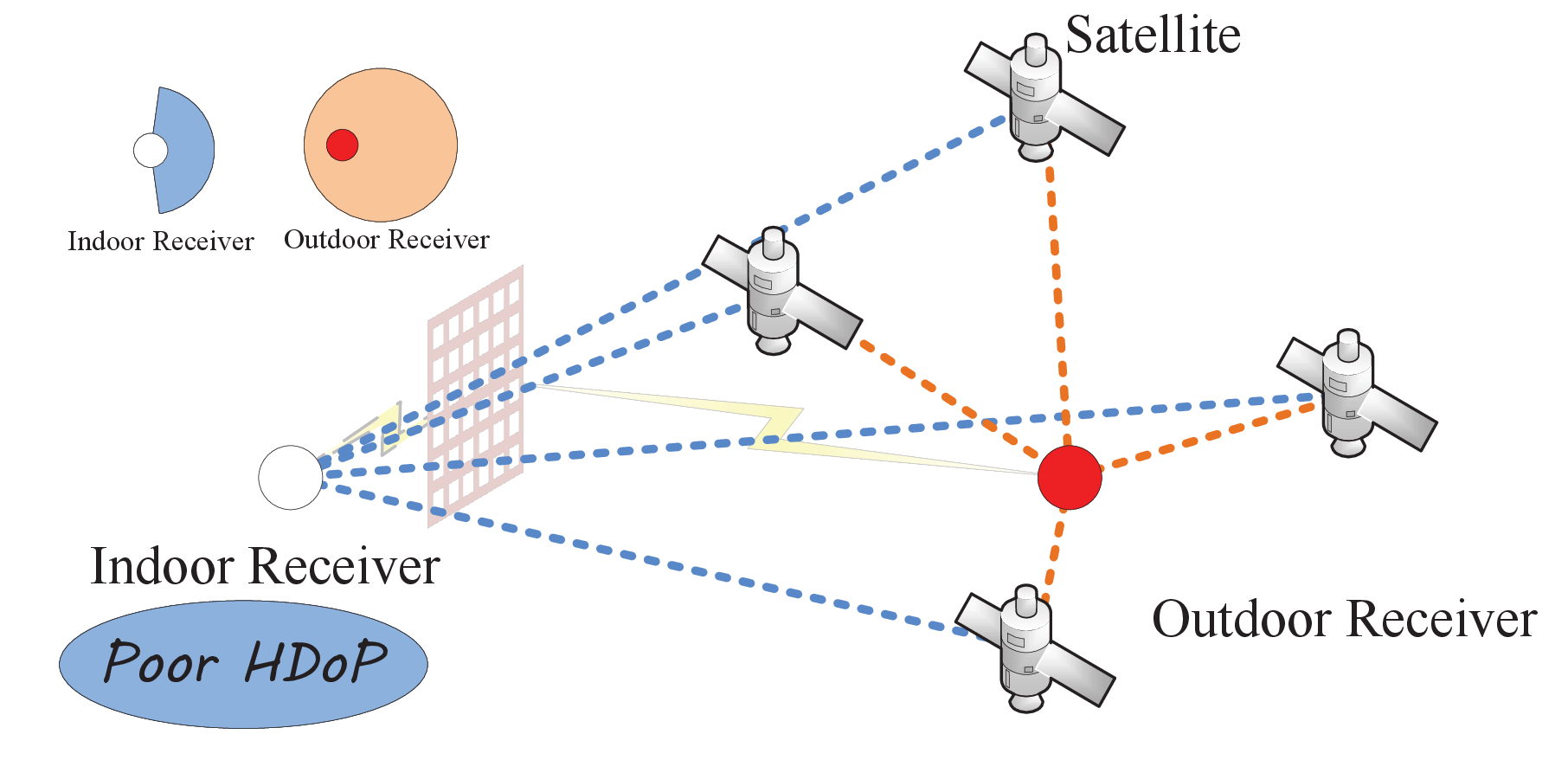}}\label{distrubution}}
  \subfigure[]{{\includegraphics[width =2.4in]{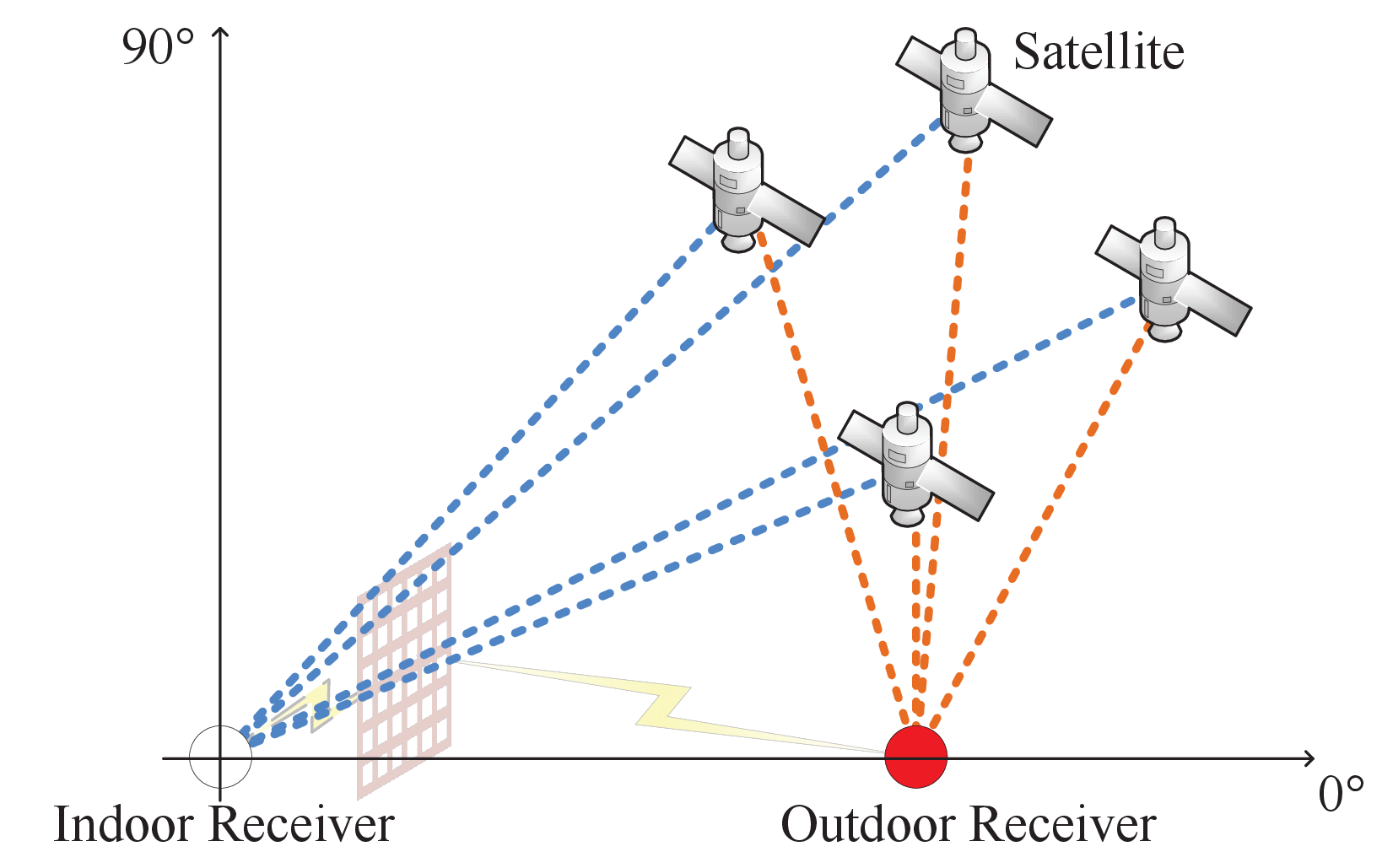}}\label{vdop}}
  \caption{Satellite azimuth and elevation angle diagram. \\(a) Satellite azimuth angle diagram. (b) Satellite elevation angle diagram.}
\label{distribution11}
\end{figure}

In the proposed ASTARS empowered satellite positioning networks, we separately discuss the impact of errors in transmission scenarios and reflection scenarios. As shown in Fig.~\ref{distrubution}, white dots represent indoor receivers, while red dots represent outdoor receivers. The fan-shaped and circular areas in the upper left corner of the schematic indicate the acceptable range of navigation signals. Fig.~\ref{vdop} shows the range of variation of the satellite elevation angle.

\begin{remark}\label{transmitRIS} In the transmit ASTARS scenario, as shown in Fig.~\ref{distrubution}, the receiver primarily receives navigation signals from satellites that are located on the opposite side of the ASTARS, which results in the azimuths of the received satellite signals being similar due to the concentration of signal sources, leading to a deterioration in the geometric solvability in the horizontal direction, which consequently increases the HDoP. At the same time, as shown in Fig.~\ref{vdop}, the elevation angles of the received navigation signals remain unaffected. \end{remark}

\begin{remark}\label{refRIS} In the reflection ASTARS scenario, the receiver receives navigation signals from the same side of the ASTARS as shown in Fig.~\ref{distrubution}. Since these signals can be distributed around the receiver in the horizontal plane, the HDoP is not significantly affected. Similar to the transmit scenario, the VDoP remains unaffected. \end{remark}

Note that a higher DoP value indicates poorer satellite geometric distribution, leading to increased positioning errors. Specifically, if the standard deviation of each satellite measurement error is denoted as $\sigma $, the standard deviation of the positioning error can be calculated as follows:
\begin{equation}\label{Xrou}
\begin{gathered}
  {\sigma _{position}} = PDoP\sigma,  \hfill \\
  {\sigma _{horizontal}} = HDoP\sigma,  \hfill \\
  {\sigma _{vertical}} = VDoP\sigma . \hfill \\ 
\end{gathered} 
\end{equation}

\subsection{\textrm{Summary of Error Analysis}}
Various sources of error in the proposed ASTARS empowered satellite positioning networks affect the positioning accuracy of the receiver, including positioning errors caused by phase shift, beamwidth, network time synchronization errors, and the impact of satellite distribution. To simplify the analysis, these errors can be combined into a comprehensive formula.

\begin{itemize}
    \item $\theta_k^A$: Phase shift generated by ASTAR in processing navigation signals.
    \item $\Delta p$: Uncertainty in receiver position caused by the finite beamwidth.
    \item $\gamma$: Time synchronization deviation due to the network delay and clock drift.
    \item $DoP$: Increased DoP values caused by the concentrated signal sources.
\end{itemize}

Considering the above errors, the total error $\omega$ (in meters) can be expressed with a comprehensive formula:
\begin{equation}\label{Summary of Error Analysis}
\omega \approx \theta_k^A{\lambda/{2\pi}} +\Delta p+c \cdot \gamma+ \sigma_{position} DoP .
\end{equation}

Based on~(\ref{Summary of Error Analysis}), we comprehensively consider the impact of various errors on positioning accuracy, providing a theoretical basis for optimizing the ASTARS empowered satellite positioning approach.

\section{\textrm{SIMULATION RESULTS}}
In this section, we present the simulation results of ASTARS empowered satellite positioning approach in urban canyons and indoor environments. In the simulation, $(-2604298.533, 4743297.217, 3364978.513)$ serves as the initial position of the receiver in the urban canyons, $(-2604398.533, 4743350.217, 3365030.513)$ serves as the initial position of the indoor receiver, with the ASTARS $(-2604348.533, 4743312.217, 3364998.513)$ deployed on a building at an elevation of $40$ m above ground level. The ASTARS can receive navigation signals transmitted by at least $12$ satellites, while the receiver is unable to receive any LoS signals from the satellites. 

Throughout the simulation, the satellite position, satellite clock bias, integer ambiguity, ionospheric delay and the tropospheric delay have been obtained in advance by model calculations. The receiver is connected to time synchronization networks for clock synchronization.
\subsection{\textrm{Error Analysis Results}}
This subsection analyzes the newly introduced errors of the proposed ASTARS empowered satellite positioning approach. 
Since the phase shift error only has a numerical effect on the carrier phase measurement, we directly perform simulation analysis in receiver position in {\bf{Subsection $C$}}.
\subsubsection{Beamwidth Error}
In the investigation of beamwidth impact on positioning error, simulations are performed for signals corresponding to three different GPS frequencies: $L1$ (1575.42 MHz), $L2$ (1227.60 MHz), and $L5$ (1176.45 MHz), with the associated wavelengths being approximately 19 cm for L1, 24 cm for L2, and 25 cm for L5.
Considering the actual dimensions of the ASTARS, the spacing among ASTARS elements is set to 0.125 m (${{\lambda}_{max}}/2$) to avoid the occurrence of grating lobes and to meet the requirements of three different navigation signal frequencies. The number of the ASTARS elements is varied from $40\times40$ to $200\times200$ . The positioning errors for different numbers of ASTARS elements are calculated to quantify the influence of element count on beamwidth.

\begin{figure}[h!]
\centering
\includegraphics[width =3.4in]{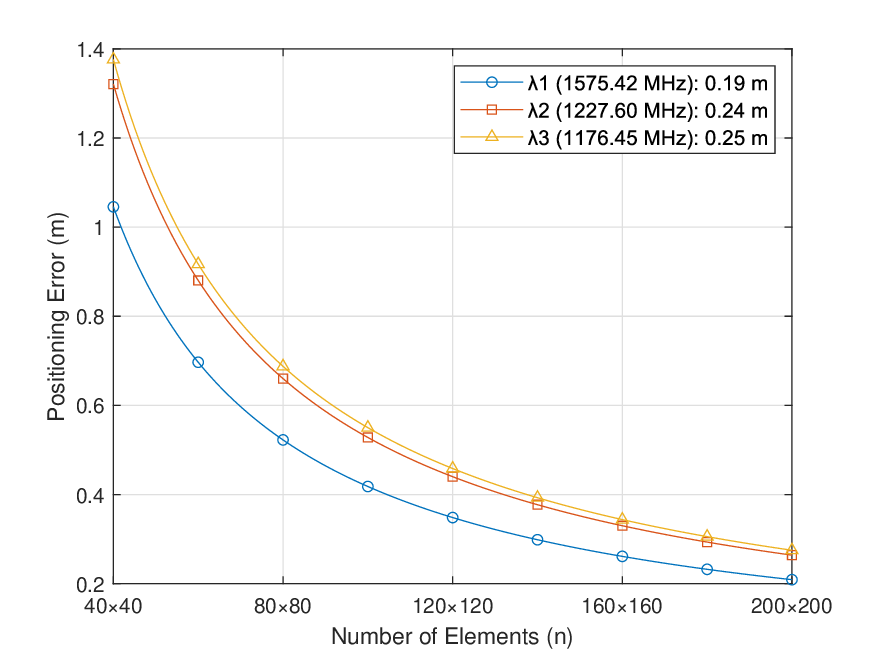}
\caption{Positioning error due to the beamwidth.}
\label{beam}
\end{figure}

As shown in Fig.\ref{beam}, the positioning error decreases progressively with an increase in the number of ASTARS elements. For the same number of ASTARS elements, higher signal frequencies (i.e., shorter wavelengths) correspond to smaller positioning errors. As the number of elements increases, the gap between different wavelengths also diminishes. When the number of ASTARS elements per row exceeds 100, the positioning error gap due to wavelength variations is within 0.1 m. Our simulation results indicate that selecting signals with shorter wavelengths, while keeping the element spacing constant, and increasing the number of elements, is beneficial for reducing the positioning error. 

\subsubsection{Network Time Synchronization Error}
The standard deviation of measurements is set to 1e-4 s, and the number of measurements per second is increased from 50 to 500. The delay variation (timing accuracy) is varied between ±100 ns.

\begin{figure}[h!]
\centering
\includegraphics[width =3.4in]{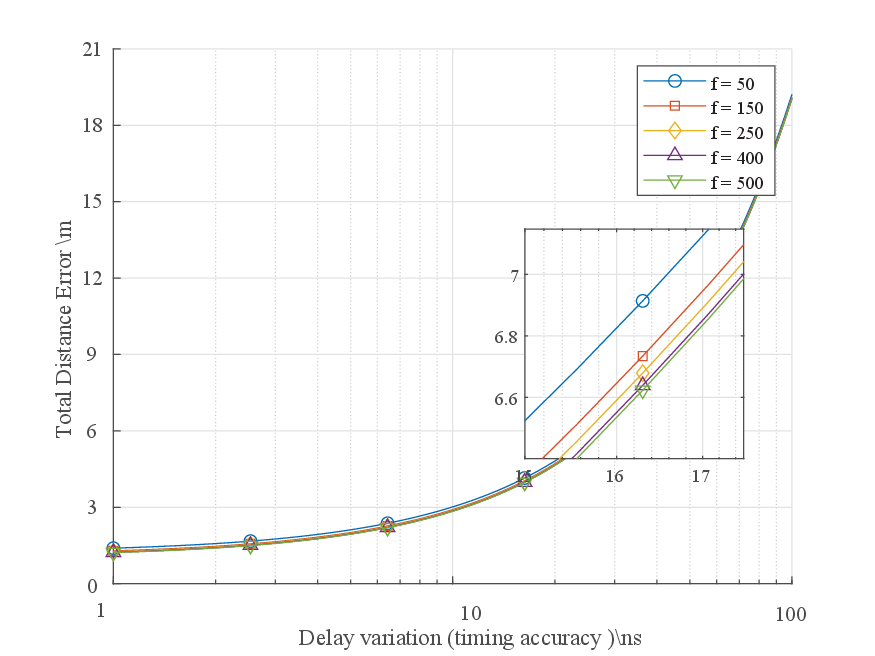}
\caption{Total distance error due to the network time synchronization.}
\label{timing88}
\end{figure}

As shown in Fig.\ref{timing88}, five different curves correspond to different numbers of measurements. It is observed that the distance error increases as the timing accuracy decreases, with minimal differences between measurements per second taken at different times. Results indicate that when timing accuracy is controlled at 10 ns, the distance error is 3 m. Therefore, providing nanosecond-level network time synchronization can keep the error caused by time synchronization within 3 m.

\subsubsection{Satellite Distribution Error}
\begin{figure}[h!]
  \centering
  \subfigure[]{{\includegraphics[width =3.5in]{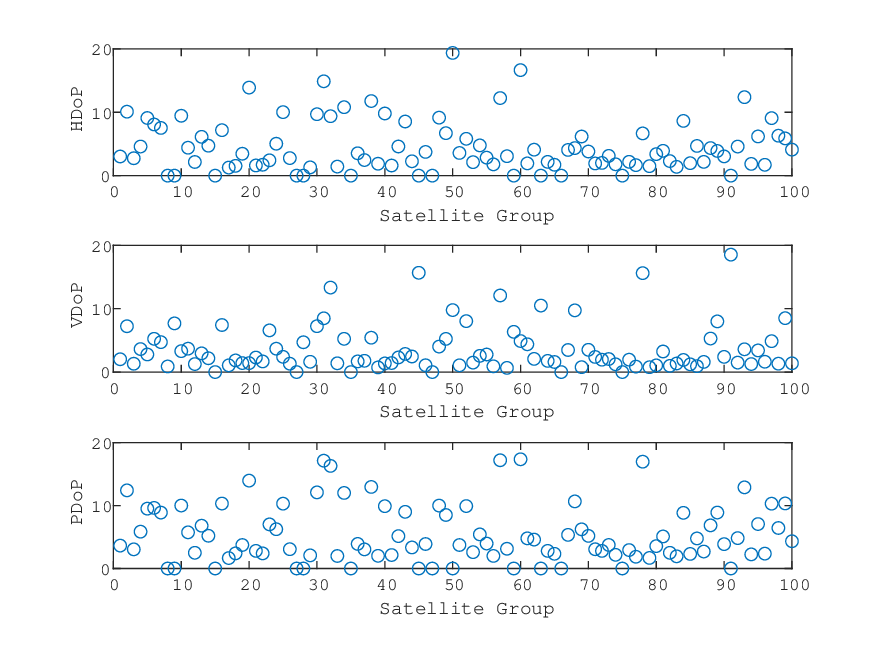}}\label{DOPtransmit}}
  \subfigure[]{{\includegraphics[width =3.5in]{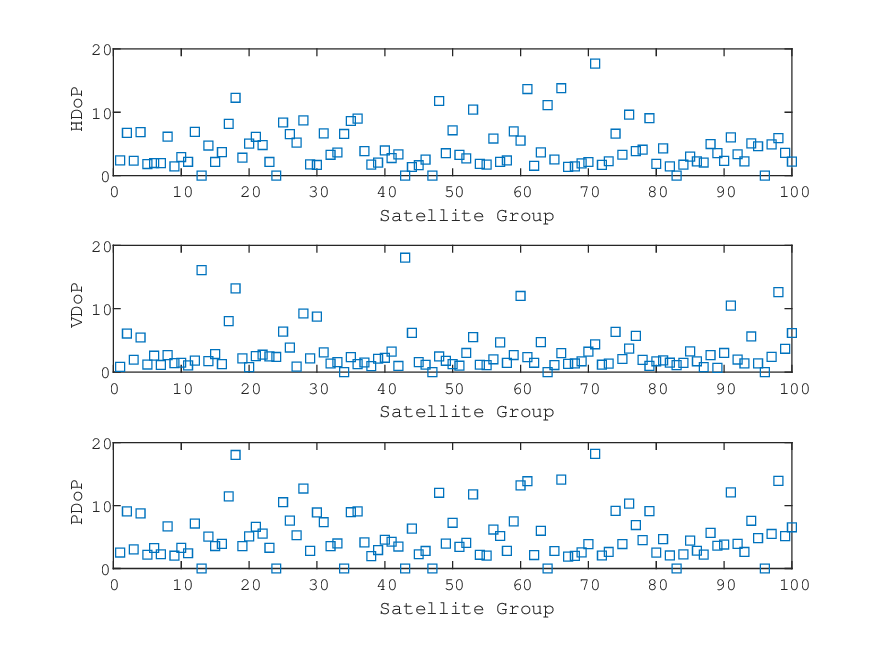}}\label{DOPref}}
  \caption{ DoP values of satellite distribution.\\(a) DoP in transmission scenarios. (b) DoP in reflction scenarios.}
\label{dop}
\end{figure}

Fig.~\ref{dop} illustrates the impact of satellite distribution on HDoP, VDoP, and PDoP. Fig.~\ref{DOPtransmit} presents the results of 100 Monte Carlo simulations for the transmission scenario, with average values of HDoP = 5.17, VDoP = 3.19, and PDoP = 6.24. Fig.~\ref{DOPref} shows the results for the reflection scenario, with average values of HDoP = 4.51, VDoP = 3.28, and PDoP = 5.72. Due to the satellite distribution, the HDoP in the transmit scenario is higher, 14.6\% greater than in the reflection scenario. In the case of the same satellite set, the elevation angle distribution in the transmission scenario is more uniform, resulting in a better VDoP value. However, the difference between the two is only 0.09, indicating that the impact of the satellite distribution on VDoP is minimal. These two conclusions are consistent with the analysis in \textbf{Remark~\ref{transmitRIS}} and \textbf{Remark~\ref{refRIS}}. Due to the higher HDoP in the transmit scenario, its PDoP value is also higher, leading to greater 3D positioning errors. 

\subsection{\textrm{Positioning Results of the ASTARS}}
In the positioning process, the number of satellites observed by the receiver through the ELoS path is typically not constant. Therefore, it is crucial to study the positioning accuracy under different satellite quantities. To this end, we tested the positioning results ranging from 4 to 12 satellites during the positioning of the ASTARS at the receiver end, comparing the accuracy under varying numbers of satellites.
\begin{figure}[h!]
\centering
\includegraphics[width =3.5in]{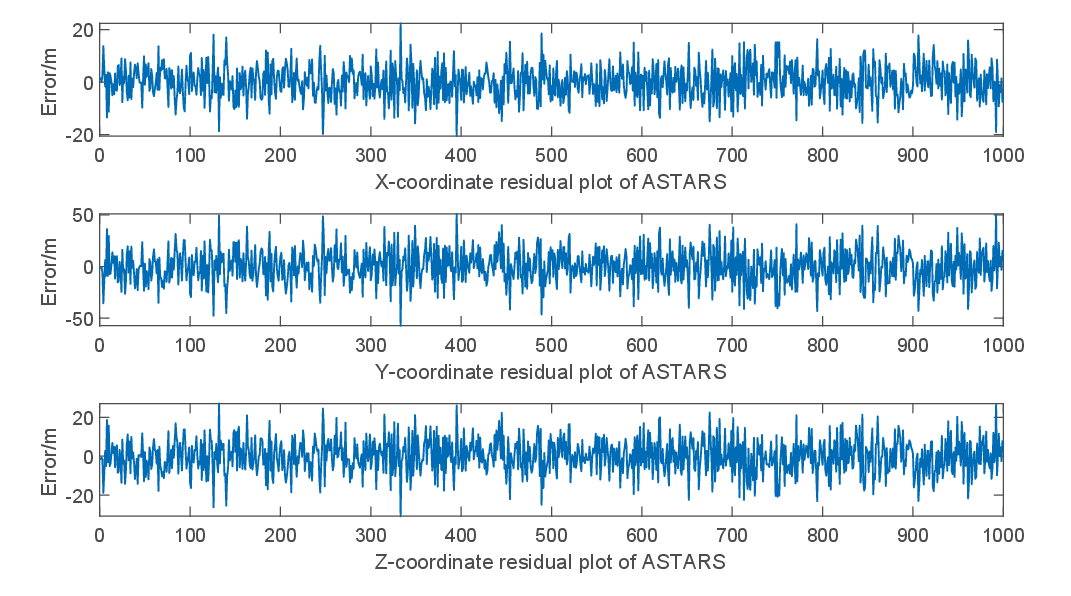}
\caption{ 3D positioning error plot of ASTARS with $N$ satellites ($N=4$).}
\label{star_ris3(4)}
\end{figure}

The results from 1000 Monte Carlo simulations, illustrated in Fig.~\ref{star_ris3(4)}, show the earth-centered earth-fixed (ECEF) positioning errors when only 4 satellites are observable. In the simulation, the errors are more pronounced on the Y-axis (horizontal direction), with fluctuations of ±20 m on the X-axis, a larger range of ±50 m on the Y-axis, and ±20 m on the Z-axis. These inaccuracies primarily result from the poor geometric distribution of satellites, as most observed signals originate from the same direction, adversely affecting positioning accuracy. Consequently, the positioning performance is suboptimal, with significant errors hindering the receiver ability to achieve precise positioning.

\begin{figure}[h!]
\centering
\includegraphics[width =3.5in]{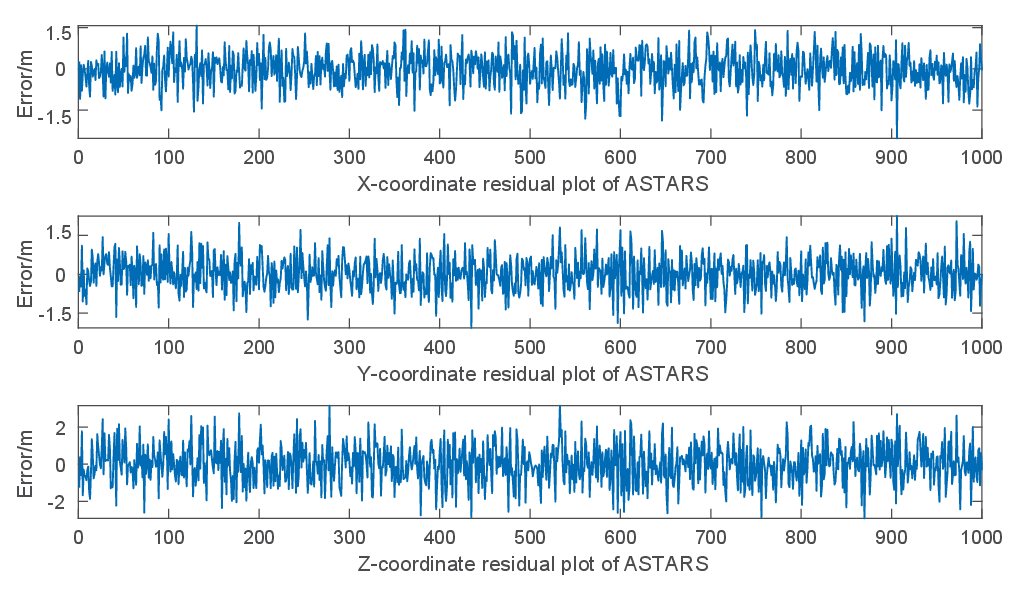}
\caption{ 3D positioning error plot of ASTARS with $N$ satellites ($N=6$).}
\label{star_ris3(6)}
\end{figure}
To address the positioning errors caused by the limited satellite distribution, we increased the number of observable satellites to 6 and conducted another 1000 Monte Carlo simulations. The corresponding 3D coordinate errors are shown in Fig.~\ref{star_ris3(6)}. 
The distribution of errors along the X, Y, and Z coordinate axes shows that the majority of positioning errors on both the X and Y axes are within 1.5 m, while Z-axis errors are contained within 2.5 m. Overall, the errors are significantly reduced, resulting in improvements in both the stability and accuracy of positioning.

\begin{figure}[h!]
\centering
\includegraphics[width =3.6in]{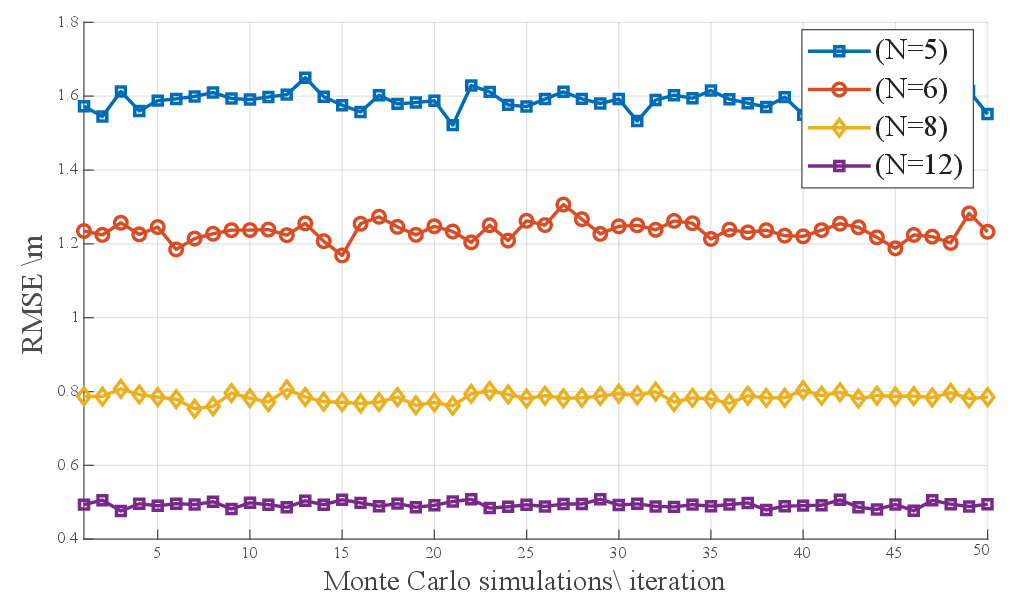}
\caption{ RMSE with the different number of satellites in ASTARS positioning.}
\label{starris_rmse}
\end{figure}
To further analyze the impact of different satellite quantities on positioning accuracy, we calculate the root mean square error (RMSE) for each iteration. As shown in Fig.~\ref{starris_rmse}, when the number of observable satellites is 5, the RMSE for the ASTARS positioning is around 1.6 m with fluctuations. However, when the number of satellites increases to 6, the RMSE decreases to 1.2 m, and the fluctuation range is reduced.
As the number of observable satellites further increases, the positioning performance continues to improve. When the number of satellites reaches 8, the RMSE drops to 0.8 m, with a fluctuation range not exceeding 0.1 m. When the number of satellites increases to 12, the RMSE further decreases to 0.5 m, with the fluctuation range narrowing to within 0.05 m.
The results show that when the number of observable satellites reaches 5 or more, the positioning performance of the ASTARS significantly improves. When the number of satellites exceeds 8, the receiver positioning accuracy for the ASTARS can achieve decimeter-level precision, indicating that the proposed ASTARS empowered satellite positioning approach exhibits exceptionally high positioning accuracy under sufficient satellite availability.

\subsection{\textrm{Positioning Results of the Receiver}}
We conduct simulations to evaluate the receiver positioning accuracy by considering the combined effects of phase shitf error, beamwidth error, network timing synchronization error, and ASTARS positioning error. In the context of practical applications of ASTARS, the signal wavelength is set to 0.19 m, the ASTARS is configured with 40 elements per row (corresponding to a length of 4 m, with an area of 16 square meters), and the standard deviation of timing errors is set to 1e-4 s, the number of measurements per second is increased to 500. On the above basis, we investigate the positioning accuracy of the receiver empowered by the ASTARS under different network time synchronization errors and the number of satellites on positioning accuracy.

\begin{figure}[h!]
\centering
\includegraphics[width =3.4in]{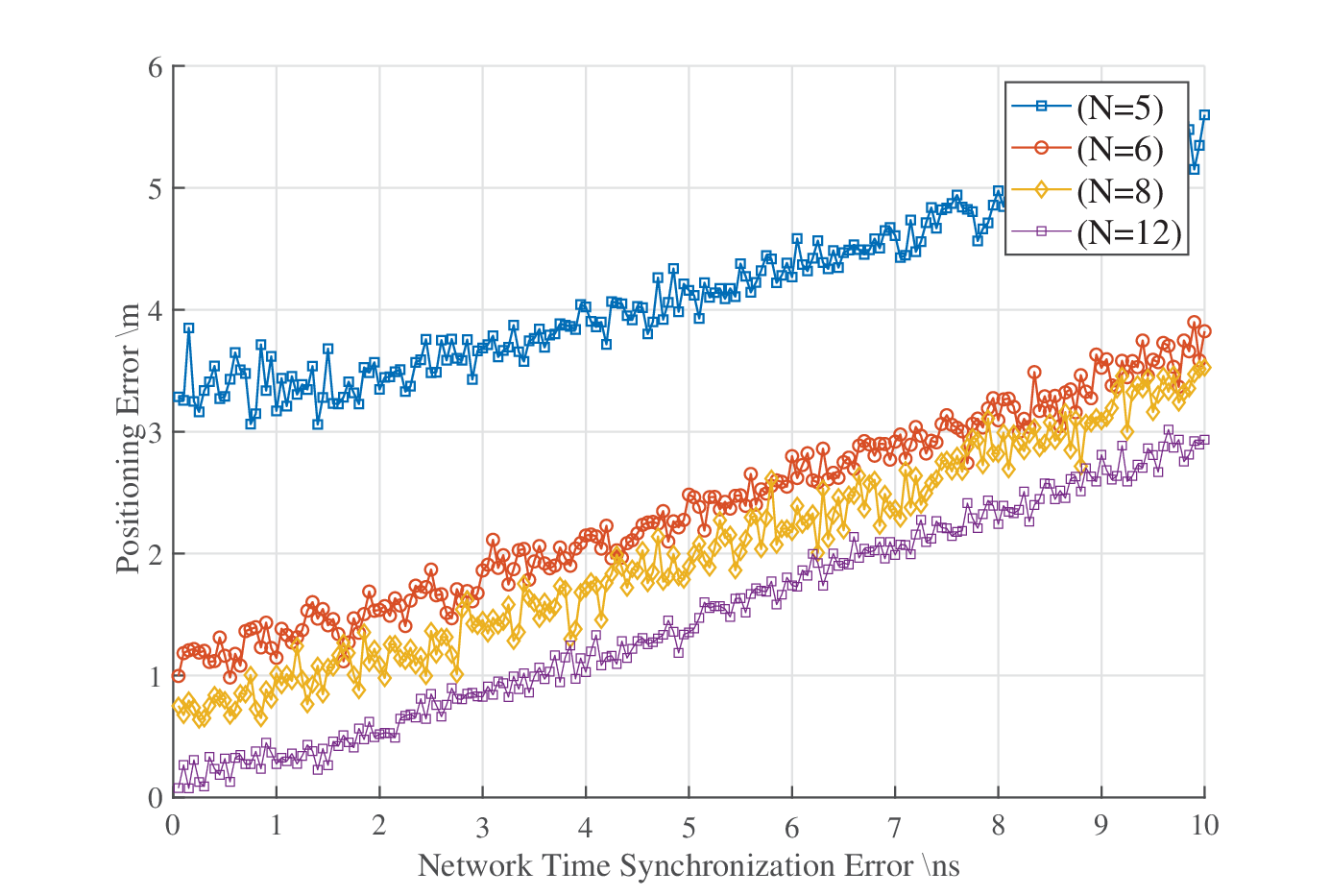}
\caption{Receiver positioning errors in different timing accuracies with $N$ satellites in urban canyons.}
\label{positioningerrorOut}
\end{figure}

\begin{figure}[h!]
\centering
\includegraphics[width =3.4in]{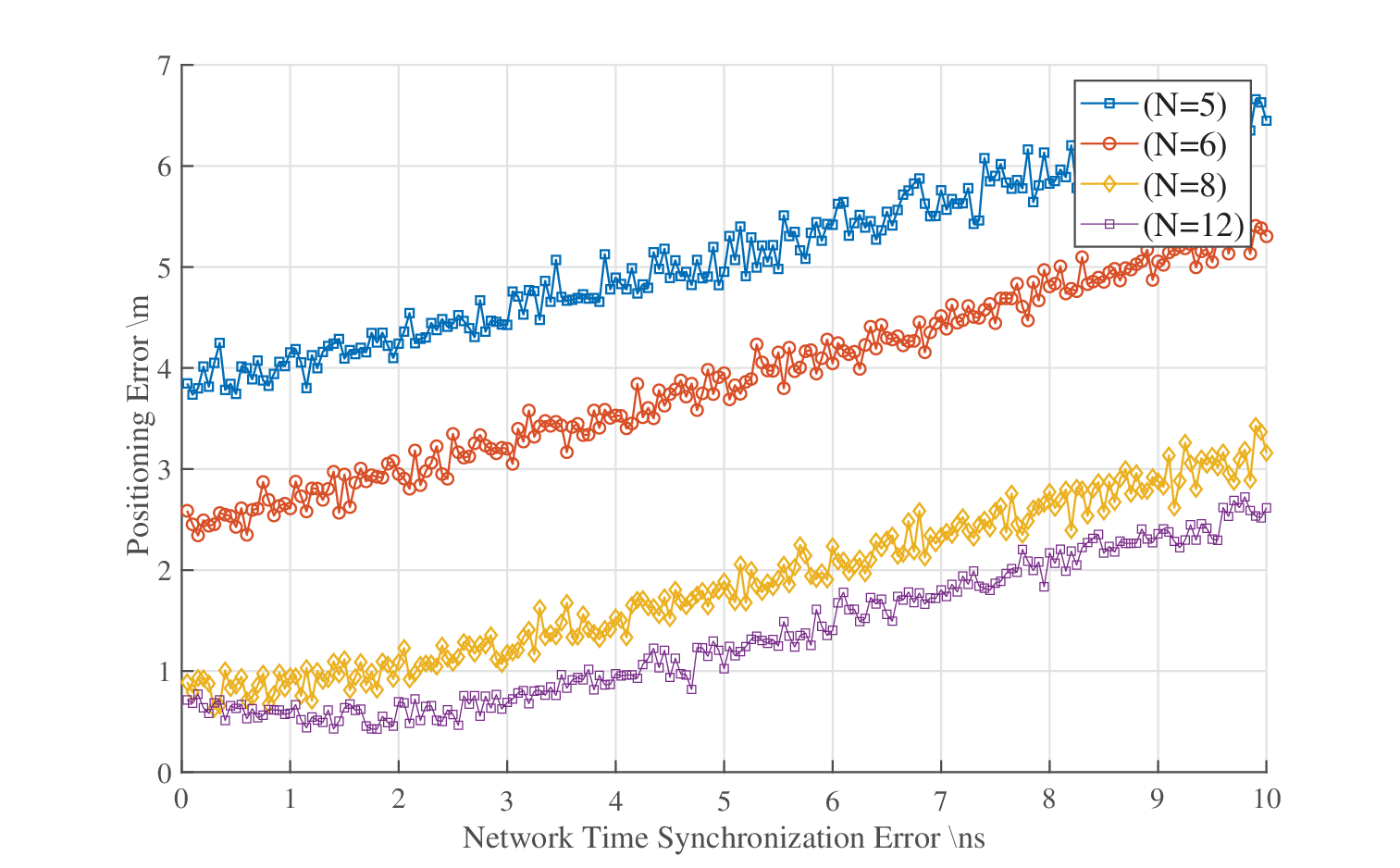}
\caption{Indoor receiver positioning errors in different timing accuracies with $N$ satellites. }
\label{positioningerrorIn}
\end{figure}
Fig.~\ref{positioningerrorOut} illustrates the receiver positioning errors under different network time synchronization precisions in urban canyons , ranging from 1 ns to 10 ns. When the number of observable satellites is 5, the positioning error corresponding to a 10 ns timing error is 5.6 m. The positioning error decreases as the network time synchronization error reduces. When the number of observable satellites increases to 6, the positioning error corresponding to a 10 ns timing error reduces to 3.9 m. With 8 observable satellites, the positioning error is between 1 and 4 m, while the number of observable satellites increases to 12, the positioning error further decreases, ranging from 0 to 3 m with minimal fluctuation. The positioning error difference between 8 and 12 satellites is approximately 0.4 meters. The results indicate that, due to the influence of beamwidth errors and ASTARS positioning errors, increasing the number of satellites to 8 or more only slightly improves positioning accuracy.

Fig.~\ref{positioningerrorIn} illustrates the indoor receiver positioning errors under different network time synchronization precisions, ranging from 1 ns to 10 ns. When the number of observable satellites is 5, the positioning error corresponding to a 10 ns timing error is 6.4 m. When the number of observable satellites increases to 6, the positioning error corresponding to a 10 ns timing error reduces to 5.3 m. With 8 observable satellites, the positioning error is between 1 and 3.4 m, while the number of observable satellites increases to 12, the positioning error further decreases, ranging from 0.7 to 2.8 m with minimal fluctuation. The positioning error difference between 8 and 12 satellites is approximately 0.3 meters. 

\subsection{\textrm{Modeled Error and Comparison of Methods }}
\textbf{Table~\ref{error-analysis}} presents the error sources in the proposed ASTARS empowered satellite positioning approach and the errors are expressed through RMSE.

\begin{table}[h!]
\centering
\caption{ Error sources in the proposed ASTARS empowered satellite positioning approach.}
\resizebox{0.5\textwidth}{!}{%
\begin{tabular}{lc}
\toprule
\textbf{Error Source} & \textbf{Typical Error Range (RMSE)} \\
\midrule
Satellite Orbit Error & 2-2.5 m \\
Satellite Clock Bias & 2-2.5 m \\
Ionospheric Delay & 2-10 m \\
Tropospheric Delay & 2-2.5 m \\
Multipath Effect & 0.1-3 m \\
Receiver Noise & 0.01-0.5 m \\
Data Processing Error & 0.1-0.2 m \\
Phase Shift Error & ${{\theta_k^A{\lambda/{2\pi}}}}$ m \\
Beamwidth Error & 0.15-0.85 m \\ 
Network Time Synchronization Error & 2-30 m \\
Total Error &  10-50m\\
\bottomrule
\end{tabular}%
}
\label{error-analysis}
\end{table}

\textbf{Table~\ref{VS}} presents a comparative analysis of five positioning technologies, including ASTARS, pseudolite, Wi-Fi, UWB, and bluetooth, across six key performance metrics: accuracy, complexity, cost, coverage, efficiency, and robustness to blockage. Among these methods, ASTARS demonstrates significant advantages in several aspects. Specifically, it achieves an accuracy of $\leq 4$ m, which is competitive with other low-cost solutions, while maintaining low system complexity and cost. Moreover, ASTARS offers broader coverage, including both indoor environments and urban canyons, where conventional GNSS signals are often unavailable. Its high efficiency and minimal sensitivity to obstructions further highlight its robustness, particularly in complex or signal-degraded environments. These characteristics collectively indicate that ASTARS is a promising solution for reliable positioning in GNSS-challenged scenarios.
\begin{table*}[t]
\centering
\caption{Comparison of positioning methods.}
\renewcommand{\arraystretch}{1.5} 
\small
\begin{tabularx}{\textwidth}{|X|X|X|X|X|X|}
\hline
Metric & ASTARS & Pseudolite & Wi-Fi & UWB & Bluetooth \\ \hline
Accuracy& \textbf{$\leq 4$ m} & $\leq 5$ m & $\leq 5$ m & 0.2–0.5 m & $\leq 3$ m \\ \hline
Complexity & \textbf{Low} & Low & Low & Moderate & Low \\ \hline
Cost & \textbf{Low} & High & Low & High & Low \\ \hline
Coverage & \textbf{Urban canyons, indoor} & Urban canyons or indoor & Indoor & Indoor & Indoor, limited outdoor \\ \hline
Efficiency & \textbf{High} & Moderate & Moderate & High & High \\ \hline
Robustness to Blockage & \textbf{Excellent}& Poor & Poor & Poor & Poor \\ \hline
\end{tabularx}
\label{VS}
\end{table*}

\section{ \textrm{CONCLUSION}}
We introduced an ASTARS empowered satellite positioning approach, offering a new positioning solution for receivers in urban canyons and indoor environments. We first reviewed the challenges faced by existing positioning techniques. Then we discussed the features and benefits of the ASTARS. Based on traditional positioning models, we developed an ELoS path empowered by ASTARS and its carrier phase observation model correspondingly. Next, we designed a ASTARS empowered satellite positioning approach, including ASTARS position calculation algorithm, ELoS path distance estimation, satellite-to-receiver distance correction algorithm, and the receiver positioning algorithm. We then analyzed the potential positioning errors introduced by phase shift, beamwidth, time synchronization and satellite distribution, evaluating its performance by using RMSE. An important future direction is to provide the ASTARS empowered satellite positioning approach for low earth orbit (LEO) satellites to address issues such as high speeds and limited coverage areas associated with LEO satellites.
\bibliographystyle{IEEEtran}
\bibliography{Enabled_by_RIS}

\end{document}